\documentclass{aa}  

\usepackage{xcolor}

\usepackage{graphicx}
\usepackage{amsmath}

\usepackage[]{hyperref}
\usepackage[normalem]{ulem}

\def\msun{\rm\, {M_\odot}}

\def\angstrom{\mathring{\mathrm{A}}}

\def\gizmo{\textsc{gizmo}}

\begin{document} 

   \titlerunning{Super-Eddington accretion in quasar hosts}
   \title{Super-Eddington accretion in high-redshift quasar hosts: Black-hole-driven outflows, galaxy quenching, and the nature of little red dots}

\author{Giada Quadri\inst{1}
        \fnmsep\thanks{g.quadri2@campus.unimib.it} 
          \and
          Alessandro Trinca
          \inst{2,3,4}
          \and
          Alessandro Lupi
          \inst{2,5,6}  
          \and
          Monica Colpi
          \inst{1,5}
          \and
          Marta Volonteri
          \inst{7}
          }

   \institute{
            Dipartimento di Fisica ``G. Occhialini'', Universit\`a degli Studi di Milano-Bicocca, Piazza della Scienza 3, I-20126 Milano, Italy
        \and
        Como Lake Center for Astrophysics, DiSAT, Universit\`a degli Studi dell'Insubria,  via Valleggio 11, I-22100, Como, Italy
        \and
        INAF/Osservatorio Astronomico di Roma, Via Frascati 33, 00040 Monte Porzio Catone, Italy
        \and
        INFN, Sezione Roma1, Dipartimento di Fisica, ``Sapienza'' Universit$\rm\grave{a}$ di Roma, Piazzale Aldo Moro 2, 00185, Roma, Italy
        \and
            INFN, Sezione di Milano-Bicocca, Piazza della Scienza 3, I-20126 Milano, Italy
        \and 
            INAF, Osservatorio Astronomico di Bologna, Via Gobetti 93/3, I-40129 Bologna, Italy
        \and
            Institut d'Astrophysique de Paris, UMR 7095, 
                CNRS and Sorbonne Universit\'{e}, 98 bis boulevard Arago, 75014 Paris, France  
      }

   \date{Received \today; accepted}

 
  \abstract{
   The advent of the James Webb Space Telescope has revolutionised our understanding of the high-redshift Universe through its detection of bright, massive galaxies up to $z\gtrsim 10$ and its identification of peculiar sources called `little red dots' (LRDs). The origin of both classes of objects remains uncertain but is likely linked to the formation and early growth of the first massive black holes (MBHs), which may be more easily explained by invoking phases of super-Eddington accretion. In this study, we used  a state-of-the-art zoom-in cosmological simulation of a quasar host to investigate whether these objects could resemble any of the peculiar sources observed with JWST during their assembly. 
   We find that the impact of MBH feedback on star formation is typically moderate, with outflows preferentially escaping perpendicular to the galactic disc. However, for approximately ten percent of the galaxy's lifetime, the system enters a distinct quenched phase following rapid MBH growth driven by super-Eddington accretion. This phase culminates in a powerful feedback event, during which the MBH jet and disc-driven winds interact directly with the galactic disc and carve out a central cavity.  We also find that, during the history of the quasar host progenitor, the spectral properties of the system can resemble both LRDs and quenched galaxies, depending on the specific evolutionary stage considered. These findings suggest that both conditions may represent transient phases in the life cycle of high-redshift galaxies.
}

   \keywords{quasars: supermassive black holes,
                accretion, accretion discs,
                galaxies: formation,
                galaxies: evolution}

   \maketitle

\section{Introduction}
Over the last few decades, observations have revealed that massive black holes (MBHs) are ubiquitously present in the Universe and reside at the centres of massive galaxies even at redshifts as high as $z \gtrsim 6$, where thousands of quasars have been discovered \citep{fan04, mortlock_11, banados18, fan23, maiolino23}. These MBHs are commonly identified during phases of active gas accretion, when the gravitational energy of infalling material is converted into radiation. Under certain conditions, the rotational energy of the MBH can also be tapped to power collimated relativistic jets. The energy released in these processes is thought to play a crucial role in shaping the evolution of the host galaxy \citep{ciotti07, fabian12, kormendy13}. For example, winds and jets can heat and ionise the surrounding gas, sweeping it away from the nuclear regions of galaxies \citep{silk98, dimatteo05} and thereby quenching star formation.
The existence of high-redshift MBHs with masses ranging from about $10^5 \ \rm \msun$ up to $10^{9-10} \ \rm \msun$ also poses tight constraints on the epoch of MBH formation and their early growth \cite[][and references therein]{volonteri21}.

To date, different mechanisms have been proposed to explain MBH formation (`seeding'): (i) light seeds, where black holes represent the remnants of Pop III stars formed inside metal-free dark matter (DM) mini-haloes (more massive than $2\times 10^5\rm\, M_\odot$), with masses ranging around $\sim 10^2 \ \msun$ \citep{madau01, schneider02}; (ii) medium-weight seeds, formed as a consequence of runaway stellar or BH mergers in very dense and compact star clusters, with masses $\sim 10^2-10^3 \ \msun$ \citep{portegies04, freitag06, lupi14, reinoso18, reinoso23, arcasedda23,rantala25}; (iii) heavy seeds descending from pristine DM  haloes with a mass around $10^{7-8} \ \msun$, in which the baryonic matter cools and collapses without experiencing significant fragmentation, resulting in seeds with a mass $\gtrsim 10^{4} \ \msun$ \citep{begelman06, omukai08, shang10, latif13, regan14, regan16, regan17, latif18, wise19, chon20, latif22}.

While all these mechanisms are, in principle, equally physically motivated and not mutually exclusive \citep{sassano21, chon25}, the formation of high-redshift MBHs faces significant constraints. A straightforward application of Soltan's argument \citep{soltan82} — which assumes that MBHs grow primarily through radiatively efficient gas accretion (with an efficiency $\eta_{\rm rad} \sim 0.1$), and under the idealised condition of continuous accretion at the Eddington limit — sets a lower bound on the initial MBH seed mass of $M_{\rm seed} \gtrsim 10^4\ \msun$ \citep{fan23}.
In reality, even when embedded in massive haloes with ample gas reservoirs, MBH growth can be significantly hindered by feedback from accretion itself \citep{milosavljevic09, dubois13}. Additionally, in low-mass galaxies, accretion can be further suppressed by stellar feedback from massive stars, which reduces the supply of cold gas in the central regions \citep{dubois15, habouzit17, lupi19}. This feedback can also cause the MBH to wander away from the galactic centre due to the shallow potential well \citep{pfister19}, thereby delaying its growth until the galaxy develops a sufficiently dense and massive bulge.

An appealing alternative, which is currently experiencing renewed interest in light of recent results from the James Webb Space Telescope (JWST), is the possibility that black holes can sustain accretion at rates that exceed the Eddington limit \citep{volonteri05, madau14, volonteri15, pezzulli16, king24, lupi24b, trinca24}, thereby compensating for their initially stunted growth.
In this respect, there is growing evidence of black holes accreting at super-Eddington rates in the local Universe, as in tidal disruption events \citep[TDEs,][]{rees88, lin17}, ultraluminous X-ray sources \citep[ULXs,][]{walton13, bachetti14}, and also MBHs \citep{du15,martinezaldama19}. One of the interesting features of this accretion regime is the weaker X-ray emission \citep{pacucci24,lambrides24,madau24}, which might explain the lack of X-ray counterparts in several recent detections of high-redshift AGN candidates.   

Further interest in these accelerated phases of growth has been generated by the detection of a peculiar class of widespread high-redshift sources identified in recent deep JWST observations, which have been dubbed `little red dots' \citep[LRDs,][]{matthee24}. These objects are characterised by red optical colours, very compact morphologies, and a distinctive `V-shaped' spectral energy distribution (SED), which presents a turnover at wavelengths around $4000\ \angstrom$ in  the source rest frame. Their peculiar SED, together with the compact size ($R_{e} \lesssim 100 \, \rm{pc}$), suggest they might be powered by a heavily obscured active galactic nucleus (AGN) surrounded by a compact proto-galaxy that dominates the UV emission \citep{hainline25}. This interpretation is further supported by the detection of broad permitted lines ($\rm FWHM > 1000 \, km/s$) in spectroscopically confirmed candidates,  which imply MBH masses in the range of $\approx 10^7 - 10^8 \, \rm{M_\odot}$ \citep{greene24,baggen24,furtak25,deugenio25}, even though alternative scenarios are still considered \citep[see e.g.][]{ananna24}. These estimates would imply (i) extreme $M_{\rm BH}/M_{\rm star}$ ratios, up to two orders of magnitude above local scaling relations, and (ii) very high number densities for bright AGNs ($\sim 10^{-5} \, \rm{Mpc}^{-3} \, \rm{mag}^{-1}$), in tension with previous estimates from UV selected quasars \citep{akins24}. However, MBH mass measurements based on the broadening of Balmer lines might be significantly overestimated for systems that accrete close to or above the Eddington limit \citep{lupi24b}. High accretion rates, which would also contribute to an intrinsically redder AGN spectrum, have been recently confirmed for at least one LRD through deep spectroscopic observations \citep{deugenio25}.

To date, only a few numerical studies have investigated super-Eddington accretion in cosmological environments and have employed  different levels of approximation \citep{dimatteo17,regan19,anglesalcazar21,rennehan24,lupi24,gordon25,husko25}. The inclusion of a physically motivated description of this accretion regime in simulations is crucial to our understanding, as it will give us a more accurate and general description of MBH evolution, in particular MBH feedback, which can be directly probed through observations. 

In this study, we further analyse the quasar host simulation from \cite{lupi24}, which includes super-Eddington accretion phases and their associated feedback in a full cosmological context. The simulation follows the evolution of a high-redshift quasar host down to $z\simeq 7.5$, and was originally presented in \citet[][L19 from hereon]{lupi19}. In particular, we explore the impact that super-Eddington accretion phases and the subsequent feedback have on the evolution of the galaxy host, with a particular focus on the quasar's outflow properties. Moreover, we assess whether these stages can reproduce some of the observational features of analogous high-redshift systems.

This paper is the fifth in a series that focuses on the characterisation of  the properties of high-redshift quasar hosts and their MBHs. In Paper I (L19), we presented and discussed the main evolution of the target galaxy and its central MBH, focusing on the stellar and gas tracers (total gas and [CII] emission), and found that super-Eddington phases were measured in the simulation, even though accretion was capped at the Eddington limit. In Paper II \citep{lupi22}, we extended the analysis and investigated the dynamics and morphology of the main galaxy as a function of redshift. In Paper III (Lupi et al. in prep.) we plan to discuss the evolution of the entire MBH population that forms during the simulation. Finally, in Paper IV \citep{lupi24}, using  the same suite of simulations employed here, we showed that super-Eddington phases might be sustained over timescales of a few tens of Myr in massive systems where large gas inflows are frequent.

The outline of the paper is as follows. In Section 2, we review the setup of our simulations and the prescriptions employed, especially the ones directly related to MBHs. In Section 3, we present our results on the importance of super-Eddington accretion phases  in the cosmological evolution of MBHs. Finally, we draw our conclusions in Section 4.

\section{Numerical setup}
This work investigates the impact of super-Eddington accretion phases on the evolution of MBHs and their host galaxies at high redshift, particularly  in massive galaxies typically observed as QSOs. In addition, we provide a detailed analysis of the resulting outflows across all accretion regimes, aiming to identify distinctive properties that could offer valuable observational constraints. To this purpose, we analysed a suite of very-high resolution cosmological zoom-in simulations evolving a rare massive halo ($M_{\rm halo}\simeq 3 \times 10^{12} \msun$ at $z \simeq 6$) down to $z \simeq 7.5$, run with the unstructured-grid cosmological hydrodynamic code \gizmo \ \citep{hopkins15}, that descends from \textsc{Gadget3} \citep{springel08} and \textsc{Gadget2} \citep{springel05}, from which the same Barnes-Hut tree solver for gravity is inherited. The simulations are performed employing the meshless-finite-mass mode, assume a gravitational softening of DM, stars, and MBHs fixed  at 40, 10, and 2.5~pc h$^{-1}$ respectively, whereas a fully adaptive gravitational softening is adopted for the gas component, whose maximum spatial resolution is of $\sim 5$~pc. In the zoom-in region, the mass resolution for baryons is of $\sim 10^4 \msun$, while the one for DM particles is around $10^{5} \msun$.
The simulation suite was performed with state-of-the-art sub-grid prescriptions that allowed us to follow in detail non-equilibrium chemistry of the most relevant species responsible for gas cooling (both primordial and metal ones), star formation, and stellar feedback, as well as MBH seeding, accretion, and feedback. In Appendix~\ref{app:methods}, we summarise the detailed description of these sub-grid prescriptions that, compared to the previous works of the series, have been refined and improved.

\section{Results}
The initial conditions were evolved down to $z\sim 9.7$  to ensure that the MBH had formed and settled near the centre of its host galaxy, but had not yet begun  to grow significantly due to the strong impact of supernova feedback \citep{dubois13,lupi19}. At this point, we split the simulation in two equivalent runs, identified as MAD and HMAD, which only differ by the magnetic flux parameter $\Phi$, either set to 1 (MAD) or 0.5 (HMAD).  The simulations were evolved down to $z\sim 7.5$ to follow the early growth of the MBH seed within its host galaxy, whose evolution will be discussed below. The results presented hereafter refer to the HMAD case.

\subsection{Global evolution of the galaxy host}
\begin{figure}
    \centering
    \includegraphics[width=\columnwidth]{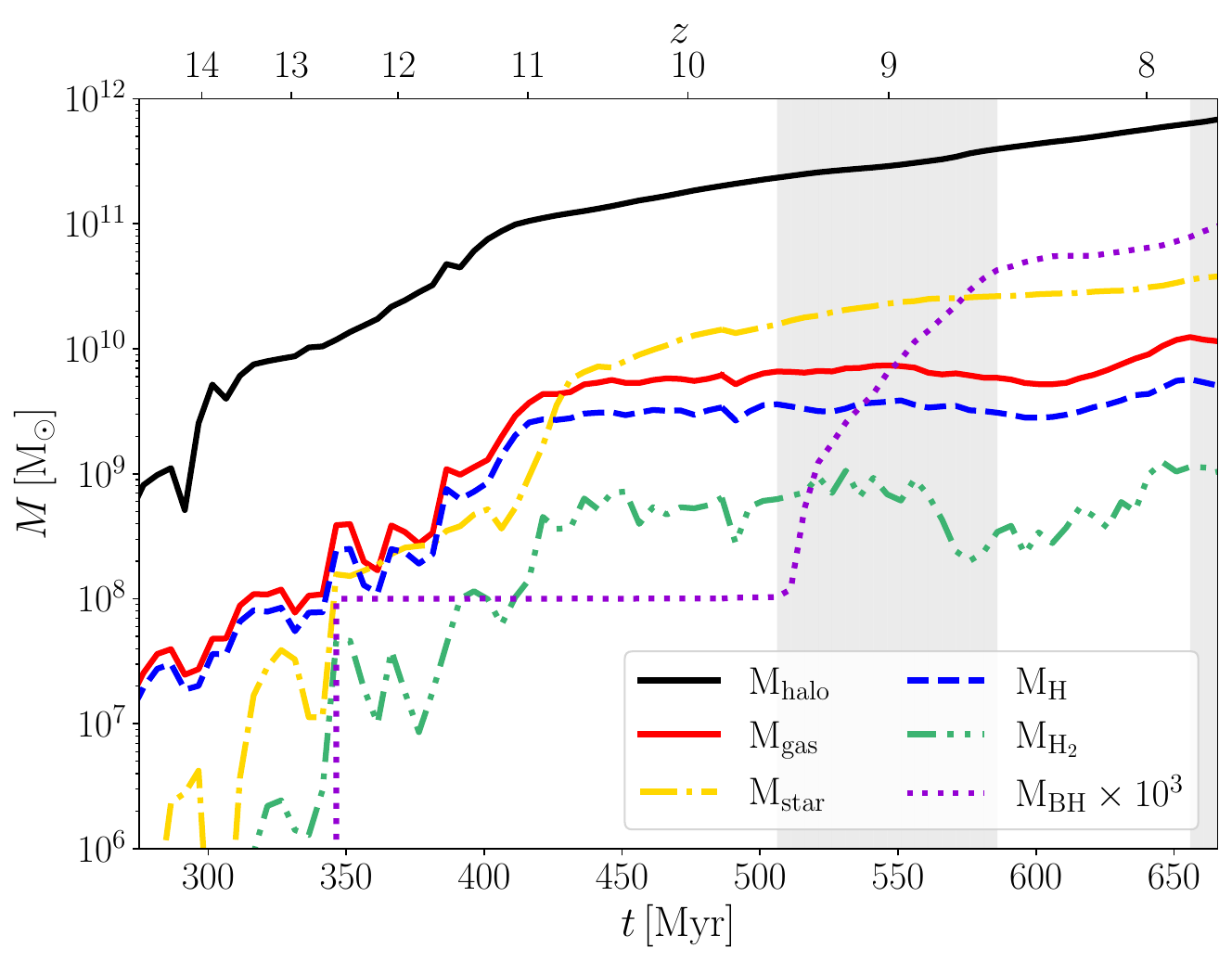} 
    \caption{Evolution of different components of the target galaxy. The solid black line corresponds to the halo mass (up to the virial radius), the dash-dotted gold one to the mass in stars, the solid red line to the total gas mass, the dashed blue line to the atomic hydrogen (H) mass content, the dash-double-dotted green line to the molecular hydrogen (H$_2$) mass content, and, finally, the dashed purple line to the black hole mass, scaled up by three orders of magnitude for visualisation purposes. With the grey shaded areas, we highlight distinct phases during the MBH accretion history: (i) super-Eddington accretion phases in the range $8.6\lesssim z\lesssim 9.5$, and (ii) accretion around the Eddington limit at $z\lesssim 7.9$.}
    \label{fig:hmad_evol}
\end{figure}
First, we discuss the evolution of the target galaxy and its host halo, identified using the Amiga Halo Finder \citep[AHF;][]{knollmann09}. In order to avoid contamination from satellite galaxies, we considered the central $20\%$ of the halo virial radius for our analysis, approximately of the order of a few kiloparsecs.

In Fig.~\ref{fig:hmad_evol}, we show the evolution of the mass of different components of the galaxy: the DM halo mass $M_{\rm halo}$, the gas mass — in particular the total ($M_{\rm gas}$), neutral ($M_{\rm H}$), and molecular ($M_{\rm H_{2}}$) content — the stellar mass $M_{\rm star}$, and the MBH mass (scaled up by a thousand for visualisation purposes). In detail, the atomic and molecular gas phases have been directly estimated from the mass fraction of the corresponding species associated to each gas particle. Consistently with the hierarchical structure formation paradigm, the mass of DM increases via mergers and inflows. Within the first $350\ \rm Myr$, the vast majority of baryonic matter is in the form of gas, because of the still low SF efficiency regulated by the strong impact of SN feedback, which efficiently heats up and sweeps gas away from the region surrounding the MBH, hindering its growth. As the galaxy mass increases, (and the potential well correspondingly deepens), stellar feedback becomes less and less efficient, reaching at $z \simeq 11$ the critical point at which the baryon content becomes dominated by the stellar component.\footnote{In L19, the mass ratio was about 1, but that was due to the inclusion of all the gas in the halo, not only that within 0.2$r_{\rm vir}$, which was used for HI, H$_2$ , and stars.}

\begin{figure}
    \centering
    \includegraphics[width=\columnwidth]{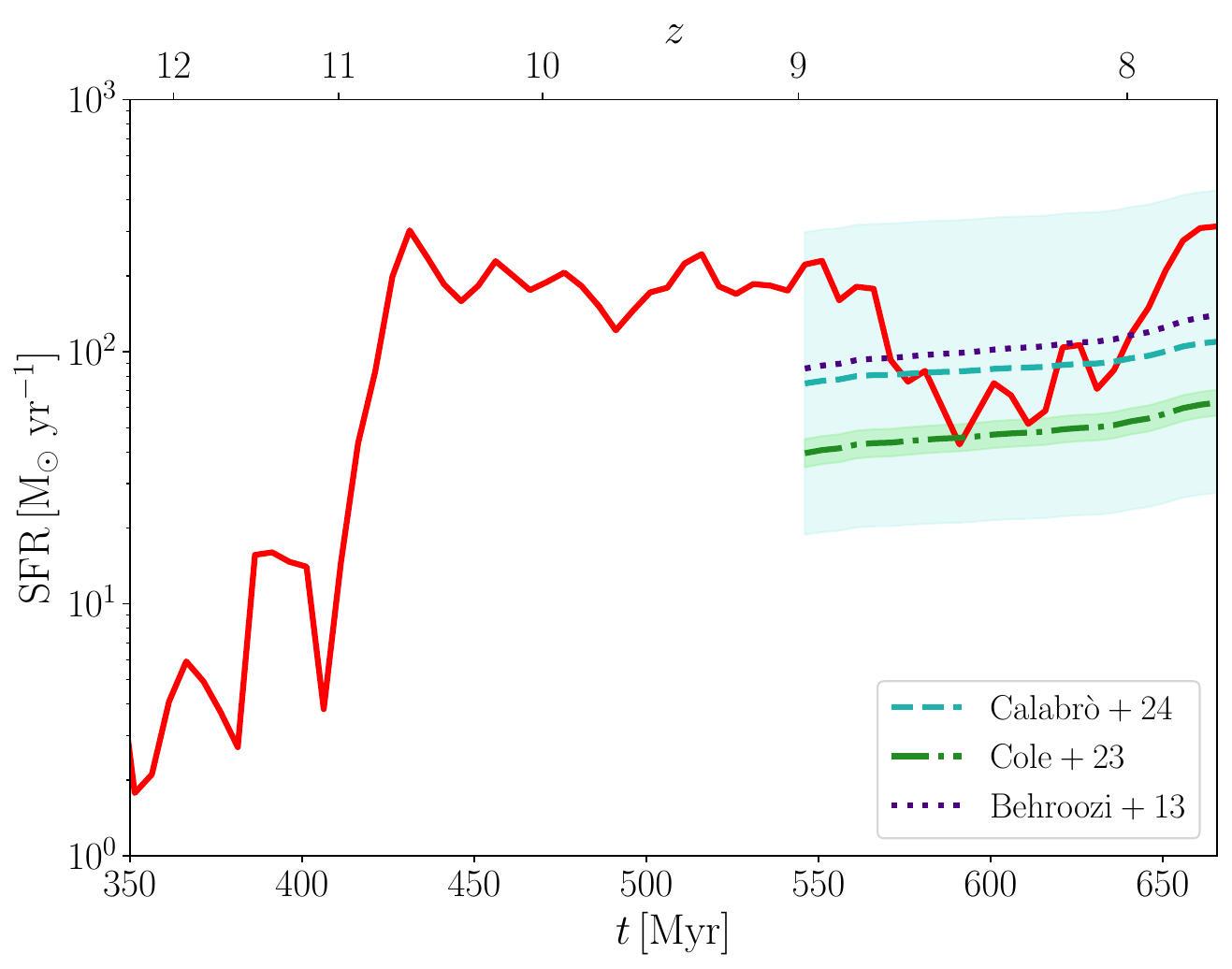} 
    \caption{Evolution of the SFR of the target galaxy (as a solid red line). The best fit of the evolution obtained from the CEERS \citep{cole23} - which explicitly excluded sources potentially contaminated by AGNs -  and the GLASS \citep{calabro24} surveys are shown below $z \simeq 9$ as dot-dashed green and dashed turquoise lines, respectively. We also report the 1$\sigma$ scatter as shaded areas, following the same colour scheme. The dotted purple line corresponds instead to the best fit to the empirical model by \cite{Behroozi13}.}
    \label{fig:SFR}
\end{figure}

In general, the galaxy evolution is similar to the original simulation of L19, but for a moderately smaller stellar mass (62\% at $z=7.8$) due to the combined effect of (i) a more accurate (and slightly less efficient) cooling at high density \citep{capelo18,lupi20b}, (ii) a slightly stronger stellar feedback \citep{lupi20}, and (iii) a more powerful MBH feedback. The MBH growth, however, differs significantly from that reported in the L19 simulation, as already noted in \cite{lupi24}. In this case, the MBH requires more time to settle at the centre of the host galaxy, and the slightly stronger stellar feedback further suppresses its growth until the galaxy reaches a stellar mass of approximately $10^{10}\rm\, \msun$. \footnote{As already mentioned in \cite{lupi24}, our simulation cannot explain the highest redshift objects. This could be due to multiple reasons, from numerical ones (a too strong MBH and stellar feedback or the use of the BHL formula) to physical ones (more favourable inflow conditions or different BH formation mechanisms).} However, at that point, the MBH starts gaining mass at a relatively high rate. In less than $50 \ \rm Myr$, the MBH mass has increased by approximately two orders of magnitude, thanks to an almost unimpeded growth at super-Eddington rates. At redshifts below $z\sim 8.7$, the MBH growth almost comes to a temporary halt. 
This outcome is driven by a phase of strong MBH feedback, which, by $z\sim 8.3$, leads to the evacuation of gas from the galaxy centre and a complete shutdown of MBH accretion. By $z\sim 8$, the cavity carved by the AGN disappears, and a new phase of accretion at the Eddington limit begins, bringing the MBH to $\sim 10^8\rm\,\msun$ by the end of the simulation. Even though the sudden halt of the MBH growth does not have any direct consequence on the galaxy dynamics, it significantly affects the abundance of molecular hydrogen, which shows a clear drop by almost an order of magnitude lasting for about 50~Myr, hence a decrease in the galaxy's SFR (as stars mainly form from cold molecular gas), as shown in Fig.~\ref{fig:SFR}.

The rapid stellar mass growth shown in Fig.~\ref{fig:hmad_evol} is driven by substantial inflows into the halo, that result in exceptionally high SFRs, as illustrated in Fig.~\ref{fig:SFR}. The SFR has been evaluated by selecting, at each snapshot, stellar particles younger than $10 \ \rm Myr$, and then estimated as the average mass formed per unit time.

As the galaxy mass increases, the impact of SN feedback weakens, reflecting in a rapid increase in SF ($z \simeq 11$), as illustrated in Fig.~\ref{fig:SFR}. This trend aligns closely with the subsequent rise in stellar mass observed in Fig.~\ref{fig:hmad_evol} and is due to a sequence of two major mergers (with mass ratio smaller than 1:2) occurring around $z \sim 11.5$. Following this initial enhancement, the SFR stabilises at approximately $\sim100-200 \ \rm M_{\odot} \ yr^{-1}$, with the notable exception of the redshift range $8\lesssim z\lesssim 9$, as discussed previously. 
For comparison, we also report recent observational results (for $z \lesssim 9$). In particular, we consider recent works by \cite{cole23} and \cite{calabro24} on two JWST-based Early Release Science (ERS) programmes, namely CEERS and GLASS. In addition, we also compare our results with the empirical model by \cite{Behroozi13}. While the overall trend recovered from our simulation is broadly consistent with the GLASS survey results by \cite{calabro24}, over the same time interval, the curve remains consistently above the evolution observed in the CEERS sample.
We note, however, that the sample by \cite{cole23} explicitly excluded sources that potentially hosted an AGN, and, as such, most likely removed sources characterised by higher SFR and larger stellar masses more in line with our simulated galaxy. This is confirmed by the results of \cite{calabro24}, in which the inclusion of potential AGN hosts led to both an increase in the best-fit relation and a larger scatter, in agreement with our simulation. Noticeably, the empirical model by \cite{Behroozi13}, based on the abundance matching technique, is in perfect agreement with the data in \cite{calabro24}, hence with our simulation as well.

\begin{figure}
    \centering
    \includegraphics[width=\columnwidth]{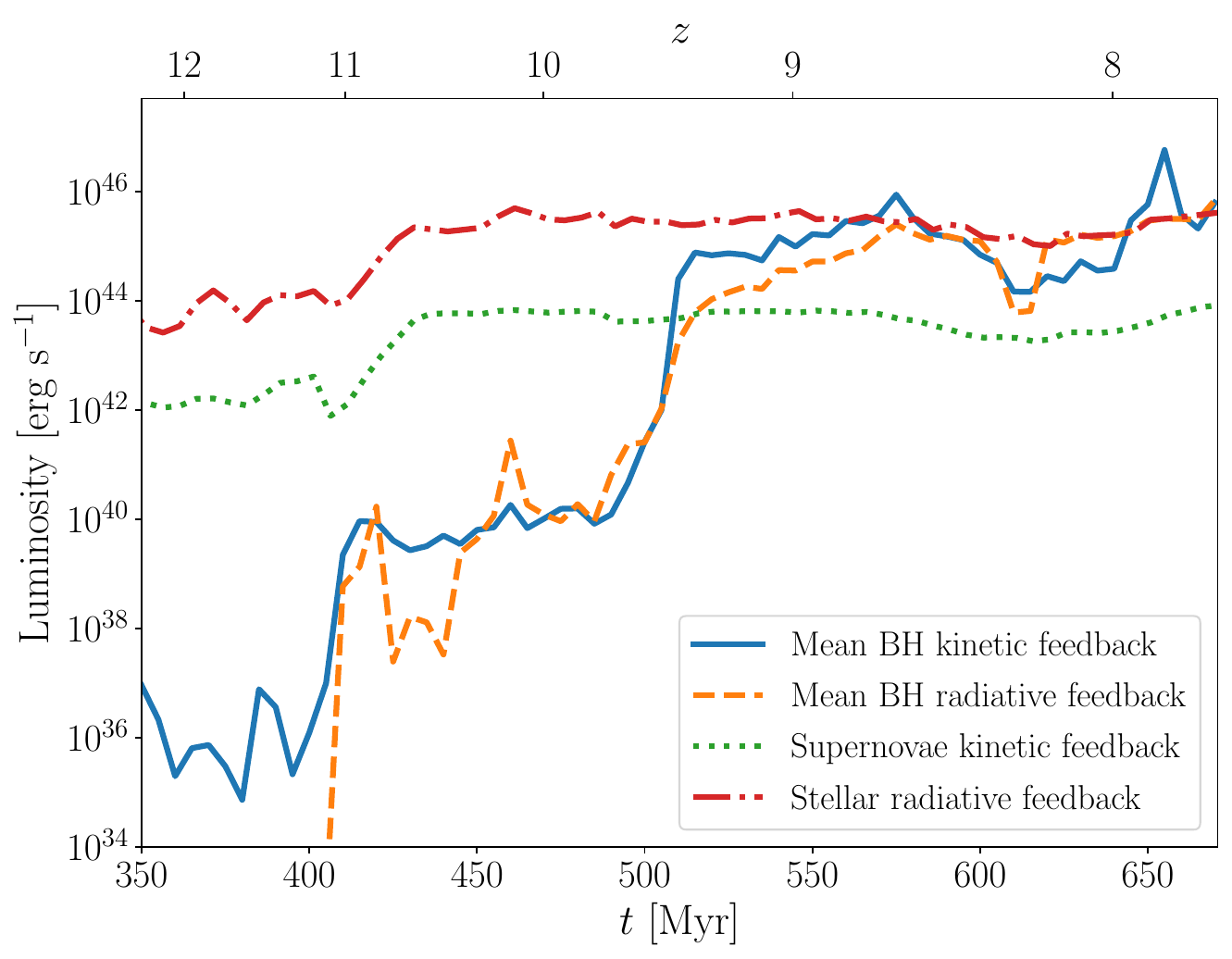} 
    \caption{Evolution of the luminosity emitted both in the form of radiation and kinetic energy by the stellar component (dot-dashed red line and dotted green line, respectively) and by the MBH accretion process (dashed orange line and solid blue line, respectively).}
    \label{fig:luminosity}
\end{figure}

In order to assess the relative role of stellar and MBH feedback on the evolution of the galaxy, in Fig.~\ref{fig:luminosity} we show the radiative and kinetic component of the luminosity emitted by stars and by the central MBH as a function of redshift. On the one hand, for the MBH we employ the prescriptions in Section~\ref{sec:feedback}, according to the accretion regime of the MBH at each time-step. 
On the other hand, for the stellar component, we consider that:
\begin{itemize}
    \item the kinetic component mainly comes from SN feedback. Starting from the equation (3) in \cite{lupi20}, the type II SN luminosity has been evaluated as
    \begin{equation}\begin{split}
        L_{\rm SNII} &= E_{\rm SN} \sum_{i}^{N_{\rm star}}\frac{ \dot{N}_{\mathrm {SN}, i}}{\rm Myr^{-1} \ {M_{\odot}}^{-1}} M_{i,0}\\ &= E_{\rm SN}\sum_{i}^{N_{\rm star}} 6.8 \times 10^{-3} \left(\frac{t_{{\rm age},i}}{t_{\rm max}} \right)^{-0.648} \frac{M_{i,0}}{t_{\rm max}}
    \end{split},\end{equation}
    where $\dot{N}_{\mathrm {SN},i}$ are the rates for type II SN, $N_{\rm star}$ is the number of stars in the galaxy, $t_{{\rm age},i}$ is the stellar particle age in Myr, $t_{\rm max}=38.1$~Myr, $E_{\rm SN}=10^{51}$~erg is the energy released by each SN, and $M_{i,0}$ is the stellar particle mass at formation.\\
    \item The radiative component has been recovered from the interpolation of luminosity tables as functions of metallicity and age. Since only photons in the UV or ionising bands directly affect the gas thermodynamics, we decided to limit the photon energy range for this analysis between $13.6$ eV and $1000$ eV, which is dominated by the youngest and brightest stars. 
\end{itemize}
For sake of clarity, instead of showing how these quantities evolve at all time-steps, we averaged the values over a time interval of $5 \, \mathrm{Myr}$, thus reducing the noise in the distribution.

In the very early stages, the MBH is in the ADAF regime, where the luminosity is almost entirely kinetic, primarily in the form of jets. This phase persists until about $z \simeq 11$, when the accretion rate increases to the point at which the MBH transitions into the sub-Eddington regime, characterised by comparable kinetic and radiative luminosities. This phase lasts until $z \simeq 9.5$, when the MBH enters the super-Eddington regime, and begins assembling mass very rapidly, as shown in Fig.~\ref{fig:hmad_evol}. Despite the strong radiative and kinetic feedback in this phase, the MBH remains in this accretion regime for tens of Myr, after which it temporarily re-enters the ADAF regime at $z \sim 8.5$, regulating its accretion, until it settles around the Eddington values, where radiative and kinetic feedbacks become almost equal. The fact that the MBH is capable of growing under this very unfavourable conditions can find an explanation in the wide variations of the evolution of the MBH accretion rate, as shown in \cite{lupi24}, suggesting that the MBH feedback on small scales, even if strong, only lasts for a limited amount of time.

Stellar feedback shows instead a very mild evolution, consistent with the almost constant SFR over the redshift range $7.8\lesssim z\lesssim 11$. By comparing the stellar radiative and kinetic feedback, we observe that radiation always dominates over SNe (by about two orders of magnitude) in terms of energy released. Nonetheless, since the effect of radiation depends on the photon cross-section, only part of this luminosity is actually translated into heating and momentum, thus making the two contributions roughly comparable.  At early times, stellar feedback dominates over MBH feedback, and this, together with the shallow potential well, explains the suppressed growth of the MBH. Around $z\sim 9.5$ instead, the MBH is massive enough and is accreting fast enough for its feedback to start dominating over that of SNe, becoming comparable to stellar radiation by $z=9$. Note, nonetheless, that despite its dominance, MBH feedback is highly concentrated in the central region of the galaxy, whereas stellar feedback is spread out across the entire galactic disc, significantly affecting the impact the two feedback sources have on the galaxy evolution. The impact of different feedback sources on inflows and outflows will be discussed in detail in Section~\ref{sec:outflows}.

\subsection{Galaxy morphology}

\begin{figure*}
    \centering
\centering
\includegraphics[height=0.90\textheight,keepaspectratio]{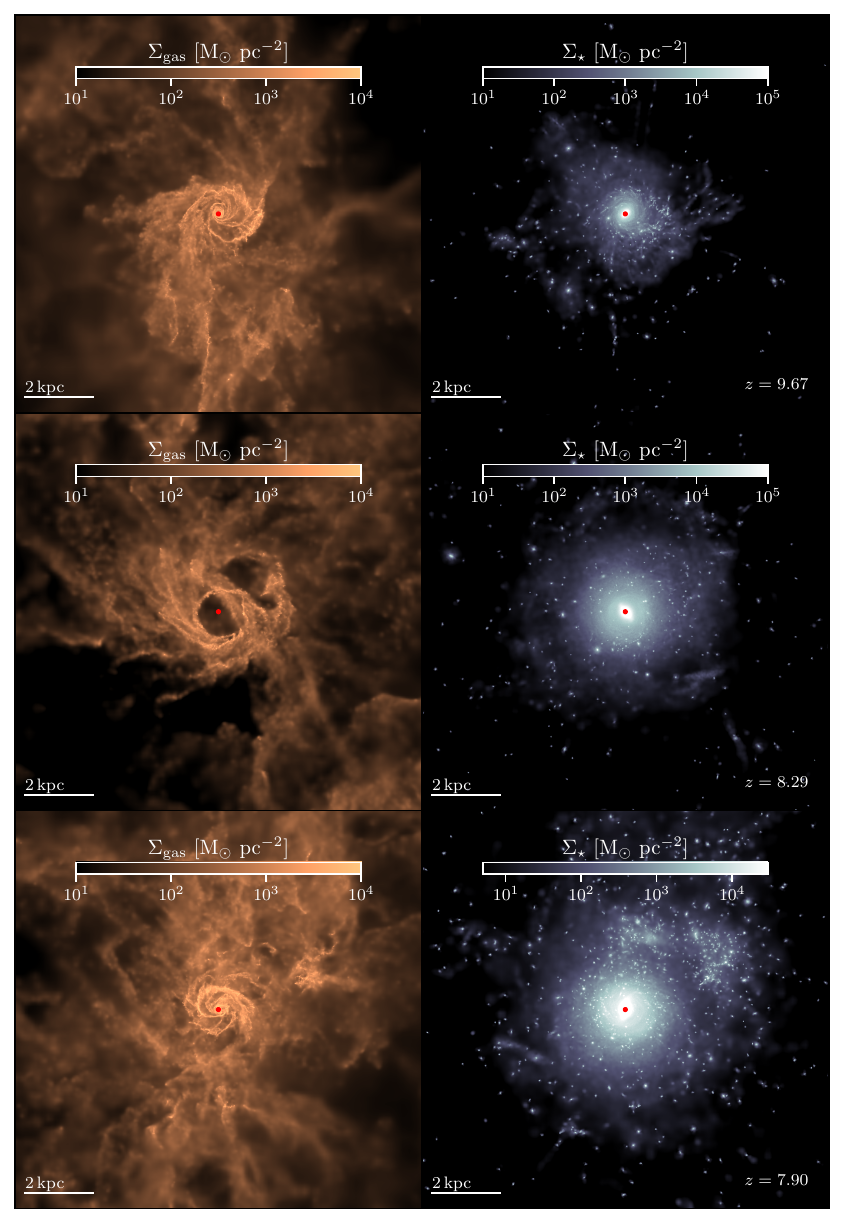} 
\caption{Gaseous and stellar column density maps in a $100$ co-moving kpc box around the target galaxy at $z \approx 9.7, 8.3, 7.9$ from left to right and from top to bottom. The middle panels show the disruption of the gaseous disc, which is absent in the stellar counterpart. The outward expanding bubble of less dense material, with its centre located at the same position of the central MBH, suggests a MBH-driven origin.}
\label{fig:density_maps}
\end{figure*}

In order to assess the connection between the integrated properties and the morphological evolution of the quasar host across cosmic time, in Fig.~\ref{fig:density_maps} we show the gas (left-hand panels) and stellar (right-hand panels) surface density distribution at $z \approx 9.7, \ 8.3, \ 7.9$. The red dots show the location of the central MBH. In general, the galaxy is characterised by a relatively compact gaseous disc, confined within a diameter of approximately 2 kpc. This disc is fuelled by larger scale filaments, which supply material to the system and influence the disc orientation by transferring angular momentum (on scales of a few hundred parsecs). However, at redshift $z \sim 8.3$, a central cavity can be clearly observed in the gas, centred at the position of the MBH. The almost symmetric distribution around the MBH suggests that the cavity has been created by MBH feedback rather than by the more distributed SN feedback. In detail, it is reasonable to assume that the jets/winds launched on pc scales (the kernel size of the MBH) shocked with the gas in the galaxy nucleus, heating it and pushing it outwards. By $z\sim 8.4$, this continuous injection of energy manifests as an expanding bubble, centred on the MBH, which starts expanding outwards, completely shutting off the MBH accretion. As the cavity expands through the galaxy interstellar medium, entraining more material, the cavity slows down, until it completely stops at $z\sim 8.1$. At later times, new  material flows in through the filaments, and pushes the gas towards the central kiloparsec, re-filling the cavity. This process rejuvenates the galaxy, recreating the gaseous disc, and reigniting SF. It is the creation of this cavity, and its survival time, which determine the duration of the `quiescent' phase of the galaxy in which the molecular gas mass, the SFR, and the MBH accretion rate (see Fig. 1 in \citealt{lupi24}, Fig.~\ref{fig:hmad_evol}, and Fig.~\ref{fig:SFR} above) 
all showed a significant drop. Notably, these drops do not exhibit the exact same duration: the amount of molecular gas is reduced around $z=8.8$, because of the increasing impact of AGN feedback, and starts rising again during the lifetime of the cavity, thanks to the new material cooling down outside the cavity; the SFR is directly affected by the reduction in molecular gas, but its increase is delayed until the cavity closes completely, because of the time needed by the gas to become sufficiently dense and lose pressure/turbulence support to ignite new SF; the MBH accretion rate, instead, remains low only during the cavity lifetime, and increases again as soon as some warm gas fills the central few pc around the MBH. 
Note that these short-time variations in the gas distribution are not uncommon at these redshifts, as already discussed in \citep{lupi22} even without super-Eddington accretion and feedback.

As for the stellar component, we find that the extreme event producing the cavity does not significantly alter the stellar distribution, but for a mild suppression of the central density, resulting from the rapid change in the underground potential previously dominated by gas. At later times, instead, a galaxy companion is found to approach the quasar host. This companion, with mass ratio around 1:10, is expected to merge with the main galaxy around $z=7.8$, as already shown in L19. Even though the companion seems gas-rich (see bottom-left panel), we expect it to be observable only by JWST, but not by ALMA \citep[see][for a discussion]{lupi22}.

\subsection{The multi-phase structure of gas inflows and outflows}
\label{sec:outflows}
It is well known that the interaction between a MBH and its galaxy host is extremely complex, and results in a multi-phase gas distribution that is detected via different tracers. As our final aim is to accurately study the (co-)evolution of galaxies and MBHs across cosmic times in detail, the inclusion of all accretion regimes in a consistent way was obviously crucial, but at the same time a self-consistent treatment of the multi-phase structure of the gas was required, that we achieved through the on-the-fly chemical solver coupled to the radiation transport. Here, we assess the impact of the MBH evolution (and associated feedback) onto the surrounding gaseous galactic disc, in particular during the super-Eddington phases which most simulations neglect.

Observationally, AGN feedback is detected in different phases, with different rates, velocities, and emission properties. The origin and evolution of these multi-phase outflows is still poorly constrained, as our understanding of the physical processes occurring close to the MBH horizon cannot be directly probed observationally, with only a few exceptions so far \citep{gillessen09,eht19,eht22,abuter24}. Moreover, the origin of the molecular phase is still unclear, with works suggesting either rapid cooling in the expanding shell \citep{richings18a,richings18b} or the lifting of molecular gas in the galaxy nucleus \citep{biernacki18}. 

\begin{figure*}
\centering
    \includegraphics[width=0.9\textwidth]{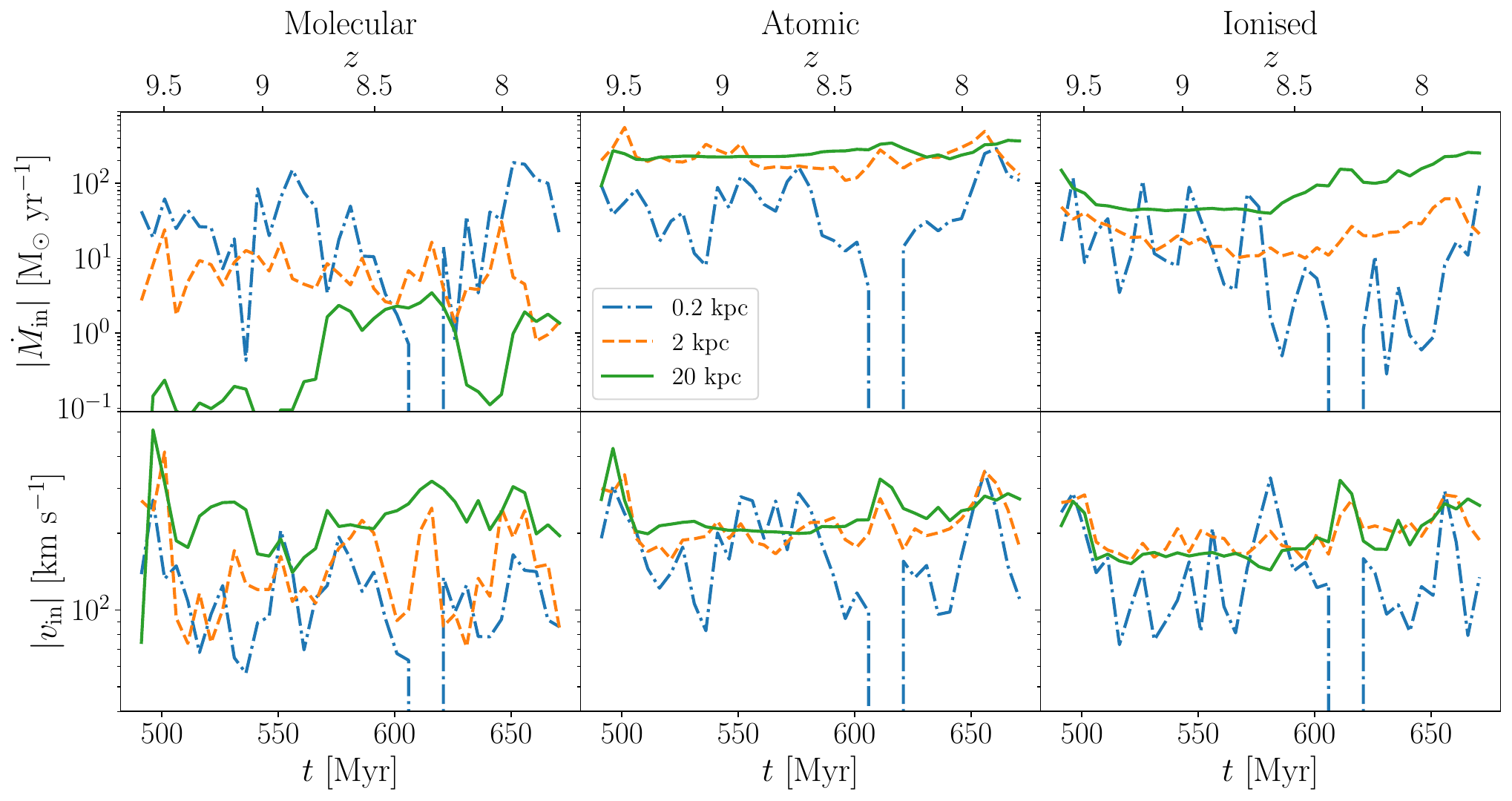} 
    \includegraphics[width=0.9\textwidth]{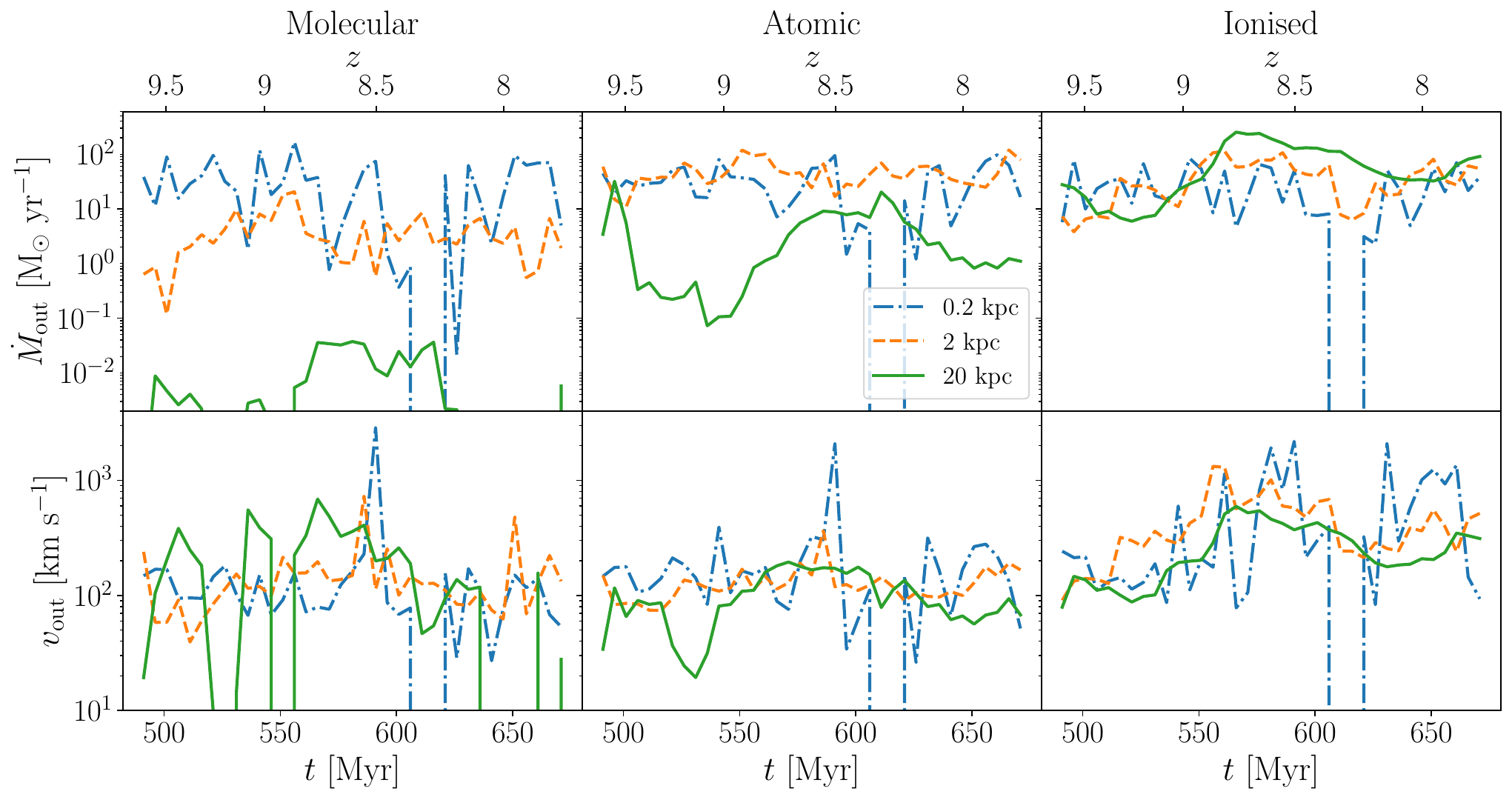} 
\caption{\textit{Top panels}: Inflow mass rates (first row) and average radial velocities (second row) for the molecular (left), atomic (middle), and ionised (right) gas phases. The solid green  lines represent inflows at $20 \ \mathrm{kpc}$, the dashed orange lines at $2 \ \mathrm{kpc}$, and the dot-dashed blue lines at $0.2 \ \mathrm{kpc}$.
\textit{Bottom panels}: Same as above but for outflows.}
\label{fig:outflows}
\end{figure*}

Thanks to our chemical network, we can constrain this multiphase structure in detail, even though our resolution is still insufficient to actually probe its origin \citep[see][for a discussion]{richings18a}. In particular, molecular hydrogen is dominant in the cold dense gas carried by inflows to the innermost region of the galactic disc, which can power star formation or end up in the central pc around the MBH. The ionised phase is instead common in the hot gas heated by feedback episodes either of stellar origin or AGN origin. Finally, neutral gas is usually found in large quantities in the outer regions of galaxies, where the density is not high enough to turn it into molecular phase, but sufficient for the gas not to become ionised, making it a perfect reservoir for later episodes of star formation. 
By looking at the relative abundance of different phases in the simulated galaxy across cosmic time, we can infer the impact of the MBH onto the evolution of the system, and also extract important information about the AGN-driven outflows.

\begin{figure*}
    \centering
\includegraphics[width=\textwidth]{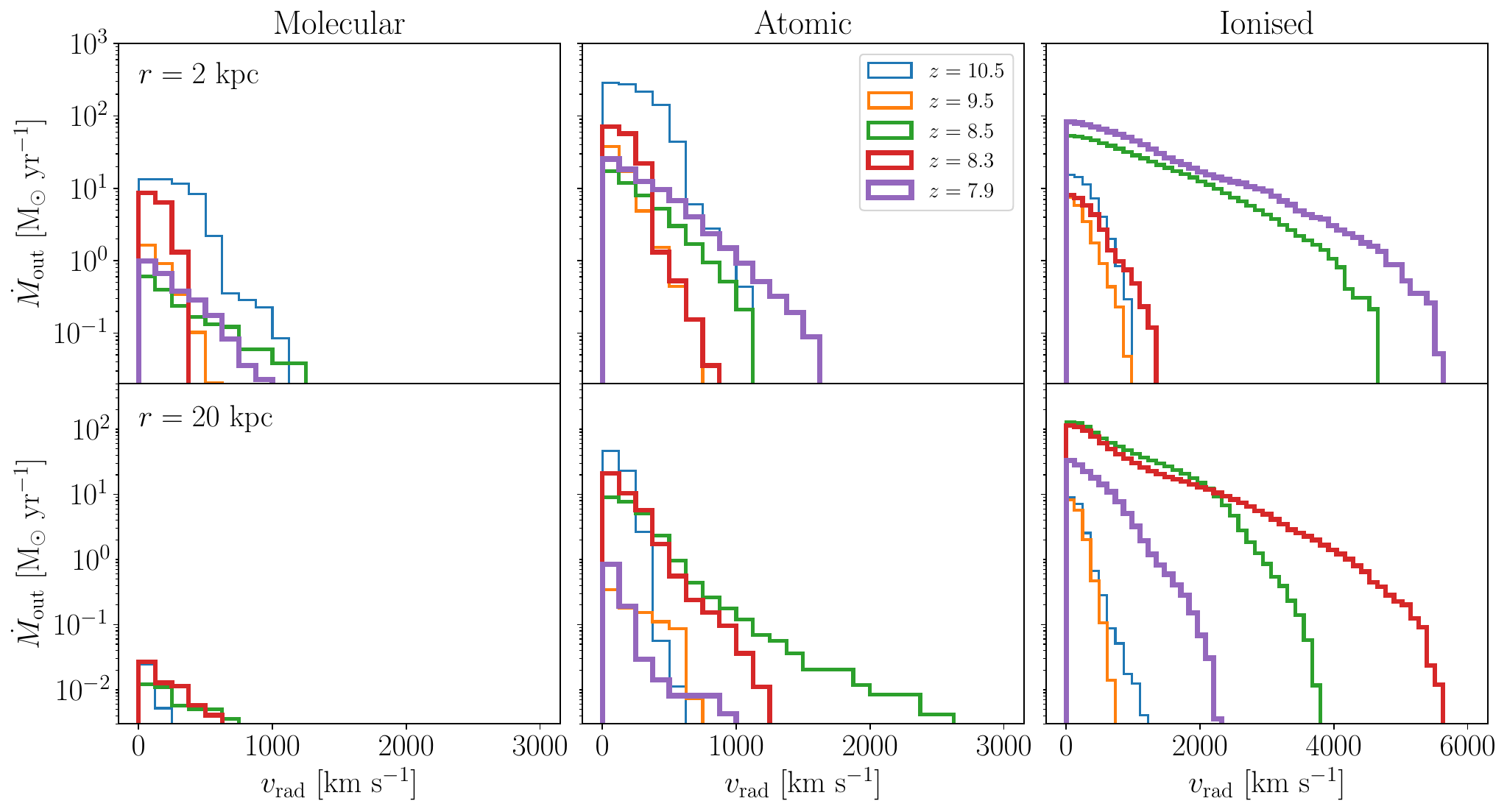} 
\caption{Cumulative velocity distribution of the mass outflows rates. From left to right, we report molecular, atomic, and ionised outflows. The rows represent spherical shells at different distances from the MBH, with the top row corresponding to $2 \ \mathrm{kpc}$ and the bottom one to $20 \ \mathrm{kpc}$. Each panel shows five distinct redshifts, with blue, orange, green, red, and purple histograms, drawn with increasing thickness, that correspond to $z=10.5,9.5,8.5,8.3,7.9,$ respectively.}
\label{fig:histograms}
\end{figure*}

Outflows driven by MBH are commonly characterised by much higher velocities, up to thousands of km/s, than those reached by stellar-driven ones.
At these velocities, the expelled gas can easily escape the galaxy (in a few Myr) and even the halo potential well, reducing the baryon fraction of the system and potentially quenching the galaxy. Note that, because of the 5~Myr output separation in our simulation, the initial expansion of these fast outflows cannot be easily reconstructed from the snapshots on a particle-by-particle basis, and can only be measured by averaging the outflowing gas around the MBH.
This is shown in Fig.~\ref{fig:outflows}, where we compare the gas inflow (top panels) and outflow (bottom panels) rates and average radial velocity at $r_{\rm shell} = 200 \ \rm pc$, $2 \ \rm kpc$, and $20 \ \rm kpc$ from the MBH, during the last 200~Myr of evolution, i.e. after the simulation split-up \citep[see][]{lupi24}. For this analysis, the rates are computed by averaging within shells of size $\Delta r = \pm 30\% \ r_{\rm shell}$, centred around each of the three values of $r_{\rm shell}$.
Note that we include in the analysis all the outflowing material, regardless of its radial velocity, hence not distinguishing between stellar and MBH-driven outflows. 

From the upper panel of Fig.~\ref{fig:outflows}, we observe that inflows, originating either from the large-scale environment or the galaxy gas reservoir, exhibit comparable velocities across all gas phases, associated to the halo/galaxy potential well. The inflow rates, instead, show a clear dependence on the distance from the MBH. At larger distances, inflows are predominantly composed of the atomic and ionised phases, as expected for low-density, warm/hot gas. Molecular gas is more abundant (but never dominant) in the central 2~kpc corresponding to the galactic disc region, where the gas is colder and denser, and more likely star-forming. Compared to the same analysis in  \citet{lupi22}, the current results show a slightly smaller molecular gas fraction, resulting from the stronger feedback (by stars and black holes), in particular in radiative form, more accurately modelled in the current simulation. 

In the bottom panels, we observe that on galactic scales ($200 \ \rm pc$ and $2 \ \rm kpc$), the outflow velocities are on average of the order of a hundred $\mathrm{km \ s^{-1}}$ in the molecular and neutral phases, whereas the ionised phase can approach a thousand of $\mathrm{km \ s^{-1}}$ below $z\sim 9$, i.e. when the MBH accretes more rapidly.  
Low velocities are typically expected to have stellar origin. Nonetheless, a non negligible contribution can also be associated to the AGN. Indeed, in our simulation winds and jets from the MBH are launched at 0.1$c$. When they shock with the surrounding ISM, the gas is almost immediately heated up above $10^4$~K, especially when the outflow propagates perpendicular to the dense galactic disc (where the denser, colder material more effectively slows down the ejected material and more rapidly cools), producing hot, ionised outflows escaping the galaxy host and propagating into the circum-galactic medium (CGM). At the same time, the post-shock gas velocity drops significantly, actively contributing to the low-velocity tail of the outflow distribution otherwise dominated by stellar feedback. In the ionised phase, the typically higher outward velocity of the gas suggests a dominant role of the MBH in driving outflows, especially after a clear path with only low-density, hot gas has been carved \citep{regan19,massonneau23}. In the neutral and molecular phases, instead, distinguishing the relative importance of the two feedback sources is harder, as they can contribute at a similar level. A clear exception is found at $z\sim 8.5$, characterised by a velocity peak of $\sim 1000\rm\, km\, s^{-1}$ within the central 2~kpc. This event precedes the formation of the central cavity discussed above and a temporary `quenched' phase of the quasar host, and is clearly associated to a massive ejective feedback phase by the MBH (hence the higher average velocity).

In terms of mass, we find $\sim 10-100\rm\, \msun\ yr^{-1}$ being ejected in the molecular and neutral phase within the central 200~pc. As the gas moves outwards (at 2~kpc), we observe a one order of magnitude reduction in the molecular component, which is compensated by a mild increase in the atomic one (due to H$_2$ being radiatively dissociated). At larger distances, the outflow rate in these phases drops, as expected given its moderate velocity, typically smaller than the escape velocity from the halo. By comparing the inflow and outflow rates, we find that the molecular phase rates are comparable, i.e. most of the molecular gas reaching the central 200~pc is expelled by the MBH. The big drop in the molecular phase at 20~kpc (apart for the MBH outburst around $z=8.5$) highlights how most of this gas falls back onto the galaxy, with the rest being dissociated as it moves outwards. In the neutral phase, inflows always dominate over outflows, as expected for the highly gas-rich environments in which the quasar host is living, continuously providing fresh cool material. For the ionised gas, the outflow rate remains within the same range across all the considered distances. This is due to two competing effects. First, the low-velocity ejected gas slows down and finally falls back onto the galaxy (assuming a momentum conserving expansion). Secondly, the neutral and molecular gas escaping the galaxy and entering the hot and turbulent CGM is heated up and ionised by the interaction with the surrounding gas and the UV radiation, increasing the ionised fraction. Additionally, the fact that the velocities remain roughly constant at different distances suggests that the relative ratio between the stellar-driven and the MBH-powered component of the outflows changes as we move away from the galaxy, with large-scale outflows being associated to the MBH feedback, but also that, at low densities, the outflow transitions to an energy-driven regime, characterised by a weaker slow down as more mass is entrained.

\subsection{Comparison with observed AGN outflows}

In order to check whether our simulated system reasonably reproduces observed sources, we now compare our results with available data. In particular, we consider both the ionised and the neutral outflows in the quasar J0923+0402 at $z \sim 6.6$ presented in \citet{bischetti24}. Using \textsc{SimBAL} spectral synthesis tool, they have identified a powerful hot ionised outflow at distances $\lesssim 210$ pc, with a mass outflow rate of $79 \ \rm{M_{\odot} \ yr^{-1}} \lesssim \rm{M_{BAL}} \lesssim 5000 \ \rm{M_{\odot} \ yr^{-1}}$ and a high velocity of $3200 \ \rm{km \ s^{-1}}$. At first sight, our outflows appear to be slower and less massive, i.e. at a similar distance the ionised gas has a velocity $v \simeq 100-1000 \ \rm{km \ s^{-1}}$ and reaches $\dot{M}_{\rm out} \simeq 5-80 \ \rm{M_{\odot}\ yr^{-1}}$. A similar comparison can be drawn for the cold neutral gas detected through ALMA observations of [CII] emission.
Note, however, that the quasar from \citet{bischetti24} is 1.5 to 2.5 orders of magnitude more massive than the simulated one, and has a much higher luminosity. 

When we compare our results with the findings of \citet{belli24}, who analysed the multiphase outflow in a massive galaxy (COSMOS-11142) at $z \sim 2.45$ instead, our velocities and the neutral outflow rate appear to be in good agreement. For the ionised phase, instead, the mass rate of the outflowing gas of our galaxy at similar distances ($\sim 3 \rm \ kpc$) is about an order of magnitude larger. Note, nonetheless, that a clear estimate of the MBH mass and accretion rate in \citet{belli24} is not present, due to the lack of X-rays, radio, and broad line emission.

In general, an important caveat in the current comparison is that the outflow rates and velocities in our simulations also include the contribution from stellar-driven ejecta, much slower than those produced by AGN. 
In order to isolate the two contributions, in Fig.~\ref{fig:histograms} we show the velocity distribution of the mass outflow rates at five distinct redshifts, for the three different gas phases. The reported redshifts have been chosen to highlight different phases in the accretion history of the central MBH: (i) accretion in the ADAF regime at $z \simeq 10.5$, (ii) the beginning of the super-Eddington growth at $z \simeq 9.5$, (iii) a typical radiatively efficient accretion below the Eddington limit at $z \simeq 8.5$, (iv) the cavity formation at $z \simeq 8.3$, and, lastly, (v) $z \simeq 7.9$ when accretion settles at about the Eddington limit. Each row represents a spherical shell, centred at $2 \ \mathrm{kpc}$ (top row) and $20 \ \mathrm{kpc}$ (bottom row) from the MBH. 

Focusing on the molecular phase, we observe that, at 2~kpc, outflowing material is present at various redshifts. In some cases, outflows reach mass rates of up to $\sim 10 \rm \  M_{\odot} \ yr^{-1}$, with velocities around hundreds of $\rm km \ s^{-1}$, supporting the hypothesis that outflows in this phase predominantly have a stellar origin. Some outflows exhibit velocities exceeding $10^3 \rm \ km \ s^{-1}$, although their mass rates are extremely low, which may indicate that the material is being accelerated. The fact that, at a distance of $20 \rm \ kpc$, the maximum outflow rate is of the order of $10^{-2} \rm \ M_{\odot} \ yr^{-1}$ further reinforces this scenario. A similar analysis can be applied to the atomic phase of the outflows at a distance of 2~kpc. Here, the outflow mass rate is of the order of $10^2 \ \rm{M_{\odot} \ yr^{-1}}$ with velocities smaller than $10^3 \ \rm{km \ s^{-1}}$. Just before and especially after the formation of the cavity (respectively the green and purple line), together with a phase in which the MBH is accreting in the ADAF regime (blue curve), we observe some high-velocity tails, although the mass rate of these outflows is relatively low, around $1 \ \rm{M_{\odot} \ yr^{-1}}$. At larger distances, a high-velocity tail reaching $\sim 0.1\rm\, M_\odot\, yr^{-1}$ is still present, suggesting a significant contribution of the AGN. In contrast, the ionised phase reveals a completely different picture. Notably, the radial velocities in this phase can reach significantly higher values. Specifically, at small distances, we observe that outflows can achieve very high velocities, above $4 \times 10^3 \ \rm{km \ s^{-1}}$, especially at $z \sim 8.5$ (green line) and $z \sim 7.9$ (purple line). At $z=8.5$, before the formation of the cavity, we find strong outflows in the ionised state, which likely heat the surrounding gas and drive its expansion outward. During the lifetime of the cavity, no ionised outflows are found, likely because of the negligible accretion rate of the MBH. Finally, when the cavity has been refilled with gas and the galactic disc has reformed, large outflows appear again, similarly to the pre-cavity phase. Interestingly, at $z=8.3$, a massive outflow is found at $20$~kpc, consistent with the expectation that the feedback event that opened the cavity has removed the gas within a few kpc, pushing it tens of kpc away from the disc with high velocities. 
 
These results allow us to improve the comparison with observed outflows. In particular, we first rescale the observed estimates to the MBH masses and luminosities of our source ($M_{\rm BH}\sim 5\times 10^7\rm\, M_\odot$ and $L_{\rm BH}\sim 10^{45}\rm\, erg\, s^{-1}$), using the scaling relations obtained for molecular and ionised winds in the local Universe ($0.1\lesssim z \lesssim 3)$ by \citet{fiore17}.
In the low-redshift sample, molecular outflow rates are of the order of hundreds of $\rm M_{\odot} \ yr^{-1}$, measured at distances between 0.1 and 3 kpc from the galactic centre. Compared to observations, our mass outflow rate is still approximately one order of magnitude lower. However, it is important to highlight that the observed molecular outflows at low-redshift are typically detected in the very central regions of galaxies, while in our case, at a distance of 2~kpc, we are already probing the CGM, outside the simulated galaxy. Additionally, our simulations have certain limitations: first, we may not have a high enough resolution to resolve the formation of molecular gas within outflows \citep{richings18a}. Secondly, by already being outside the galaxy, the gas can mix with the hot gas in the halo, and is also exposed to the dissociating UV background, which may prevent molecule formation altogether. As for ionised winds instead, the radii at which they are typically observed range from sub-kiloparsec scales up to tens of kiloparsecs, which are comparable to the distances probed in our simulation. In this case, our results are more consistent with the observations: for a galaxy with the same bolometric luminosity as in our simulation, we find comparable mass outflow rates of the order of $1-100 \ \rm M_{\odot} \ yr^{-1}$, except around the time of the feedback-driven cavity formation, where we expect enhanced outflow activity.

The results just described clearly show how the MBH feedback in our simulation is able to produce extremely fast outflows, that have nonetheless a moderate impact on both the MBH growth and the star formation in the galaxy. \citet{lupi24} showed that the MBH accretion rate exhibited extremely wide variations during super-Eddington phases, suggesting that the impact of MBH feedback on small scales was relatively strong, but only for a limited amount of time, yielding an average growth above the Eddington limit.

\subsection{Cavity opening and galaxy quenching}
Figure~\ref{fig:density_maps} and Fig.~\ref{fig:outflows} show that, at one point during the simulated MBH's lifetime, a particularly catastrophic event occurred, resulting in the formation of a cavity within the central 200~pc of the galaxy. To better understand the role of the MBH in creating this cavity — and to evaluate the potential observational implications of such events — we now focus our attention on the redshift interval around $z\sim 8.3$. 

\begin{figure*}
    \centering
        \includegraphics[height=0.7\textheight]{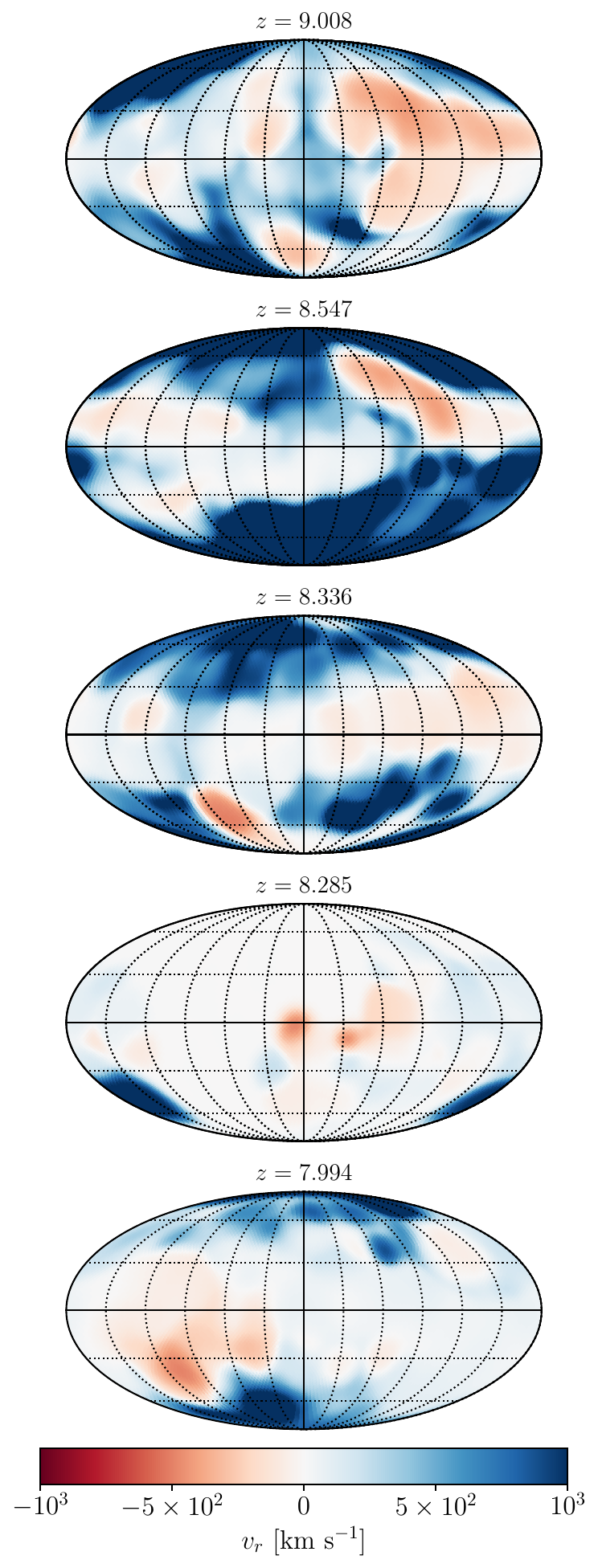}
        \includegraphics[height=0.7\textheight]{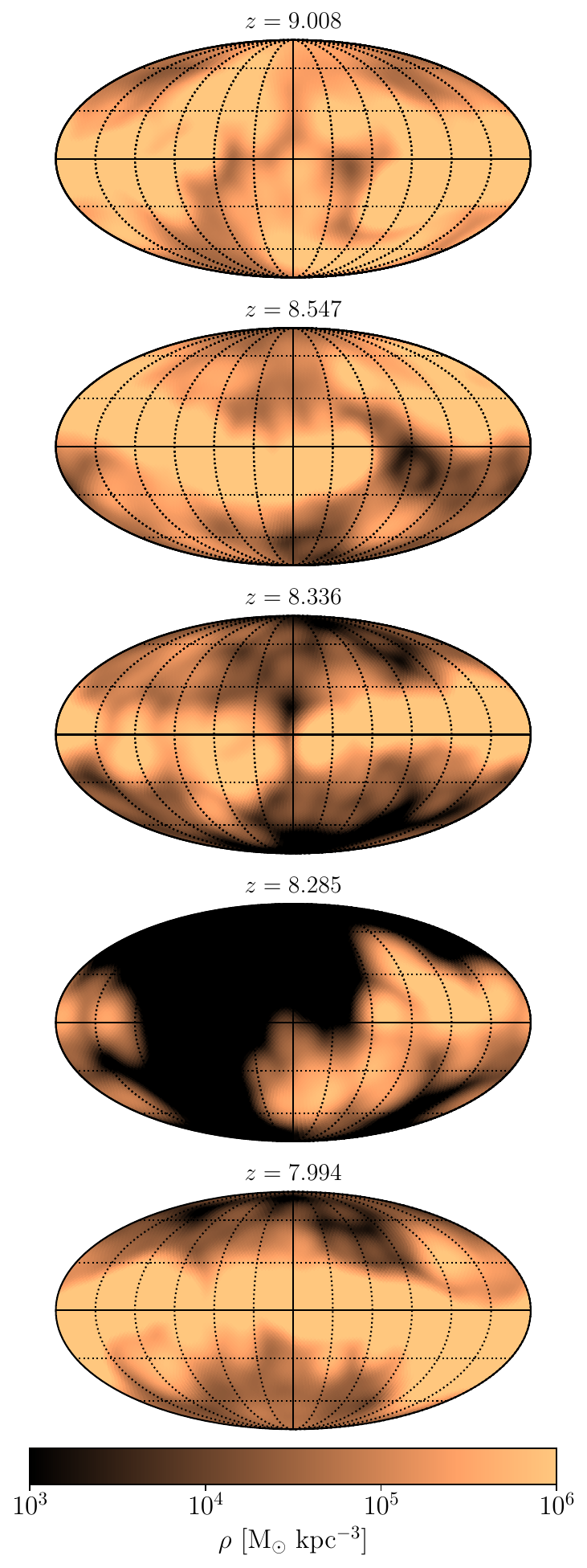}
    \caption{Radial velocity (left-hand panels) and density (right-hand panels) projection of the gas particles within a spherical region with a radius of $500 \ \rm pc$, assuming the equatorial plane aligned with the stellar galactic disc. We report five different redshifts, from top to bottom, two before the cavity creation, one when the cavity is at its maximum extent, and one after the cavity closes. The entire set spans a time interval of about 100~Myr.}
    \label{fig:healpy_maps}
\end{figure*}

First, we examine how both inflows and outflows are spatially distributed with respect to the disc of the galaxy. Fig.~\ref{fig:healpy_maps} shows the average radial velocity (left column) and the column density (right column) maps of gas particles contained within a sphere of radius $500 \ \mathrm{pc}$ centred on the MBH. The equatorial plane in all the projections is aligned with the stellar galactic disc. At early times (top panel), the MBH is embedded in high-density gas, and the AGN outflows mostly escape perpendicular to the galactic disc, resulting in an almost continuous super-Eddington growth. Around $z=8.5$, AGN feedback at more than 1000$\rm\, km\ s^{-1}$ starts to clear out the gas around the MBH (second row), reducing the average accretion rate to sub-Eddington values. The galactic disc that is forming appears perturbed by the AGN outflows, and not well aligned with the equatorial plane. In the third and fourth rows, we see the effect the AGN outburst that started around $z=8.7$ has on the galactic disc: a patchy disc-like distribution is initially created (third row) and the disc is then completely destroyed (fourth row). During this phase, inflows almost vanish from the region. At later times (last row), the cavity is replenished by new gas coming from larger scales, recreating the disc (although with still strong residual perturbations).

\begin{figure*}
    \centering
    \begin{tabular}{c c}
            \includegraphics[width=0.45\textwidth]{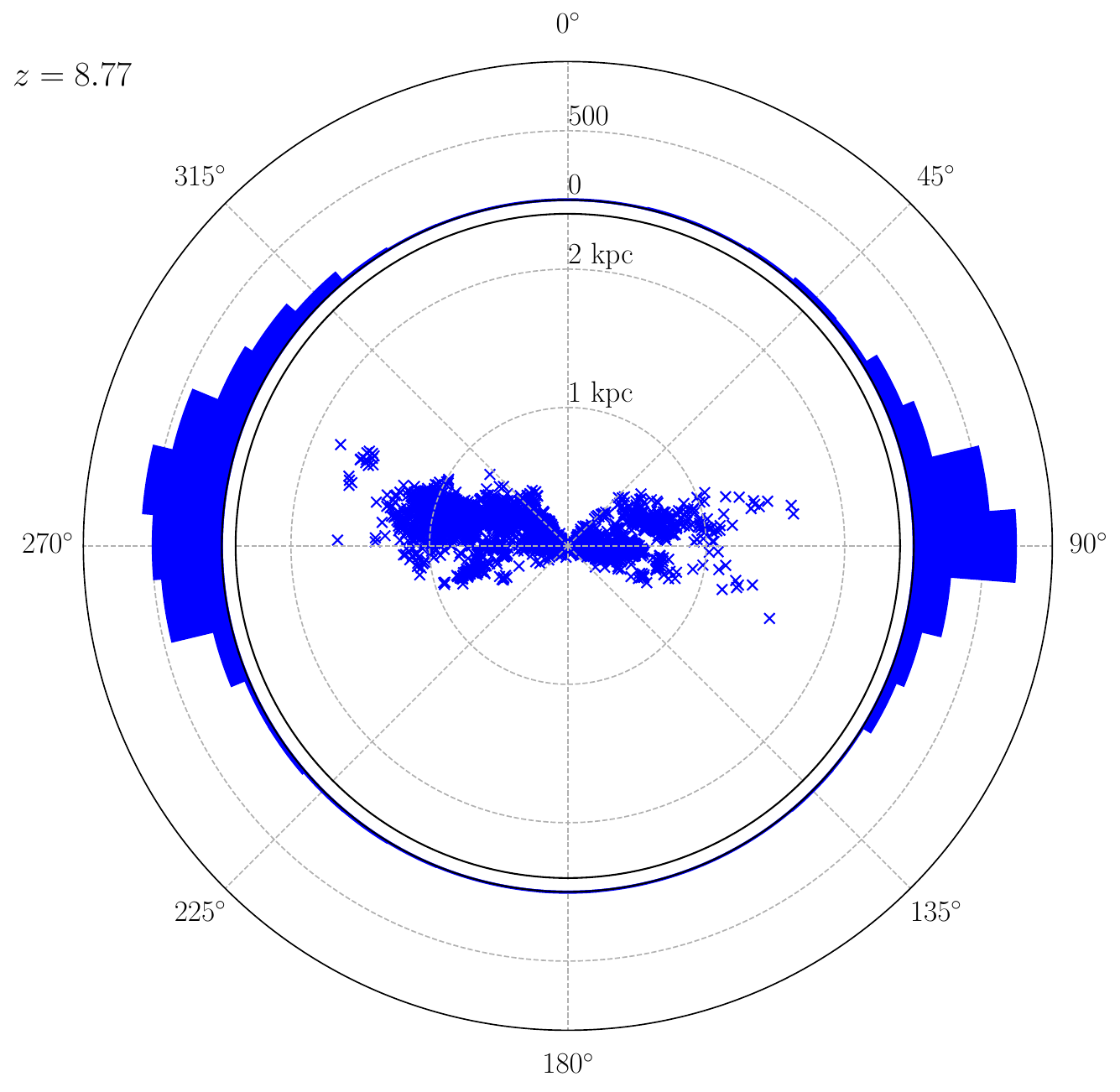}
         &  
            \includegraphics[width=0.45\textwidth]{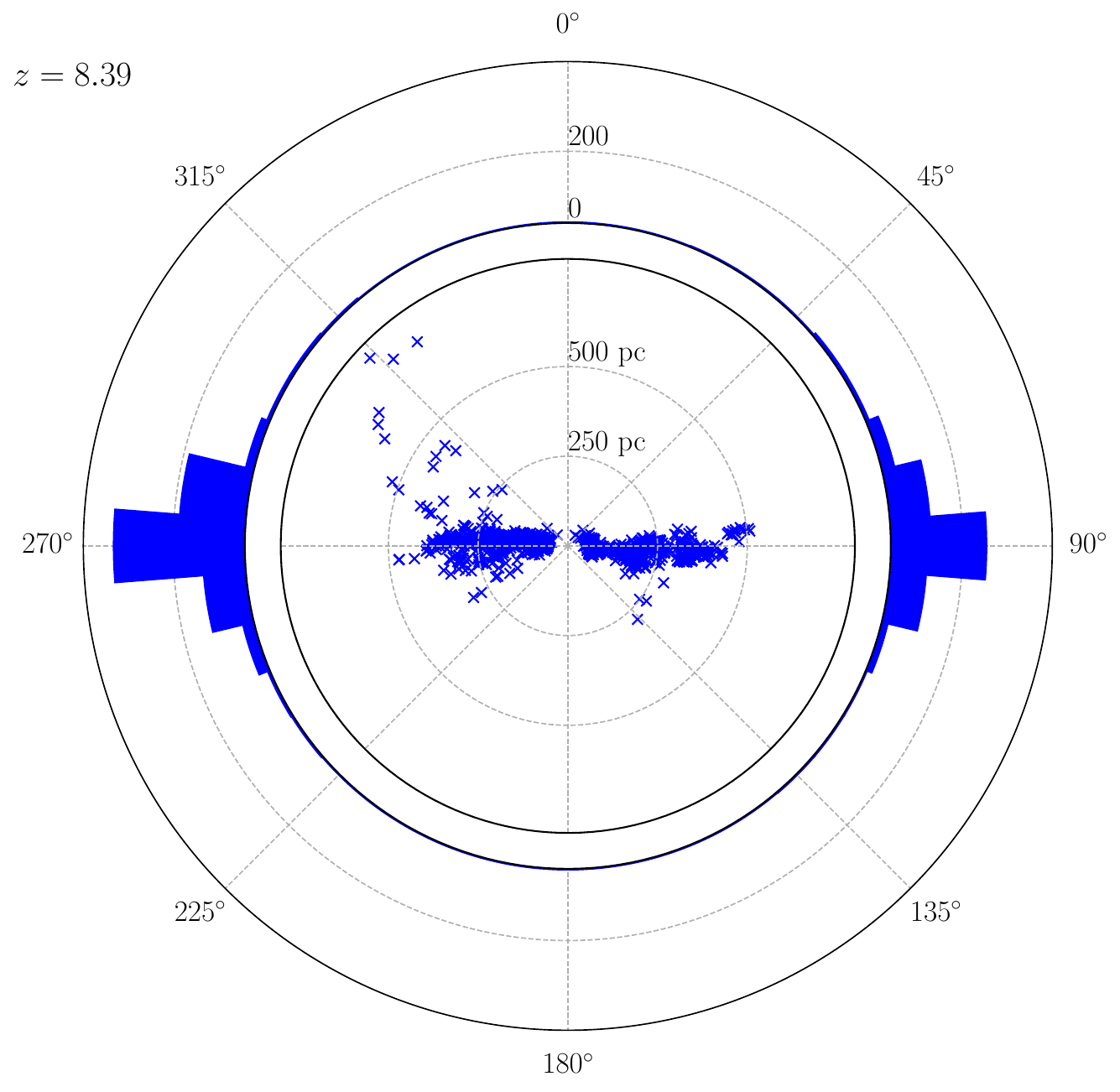}
         \\
            \includegraphics[width=0.45\textwidth]{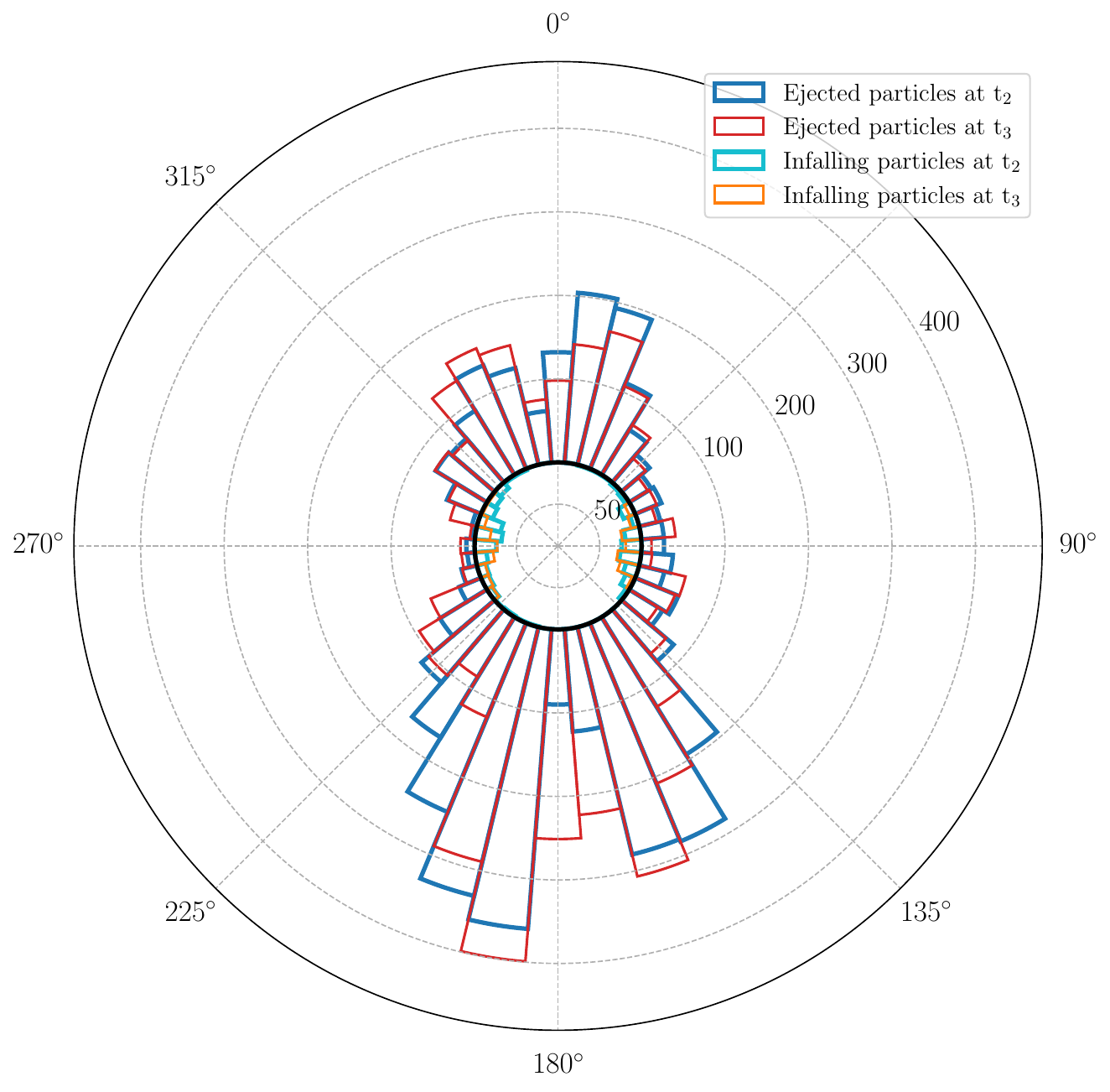}
         & 
            \includegraphics[width=0.45\textwidth]{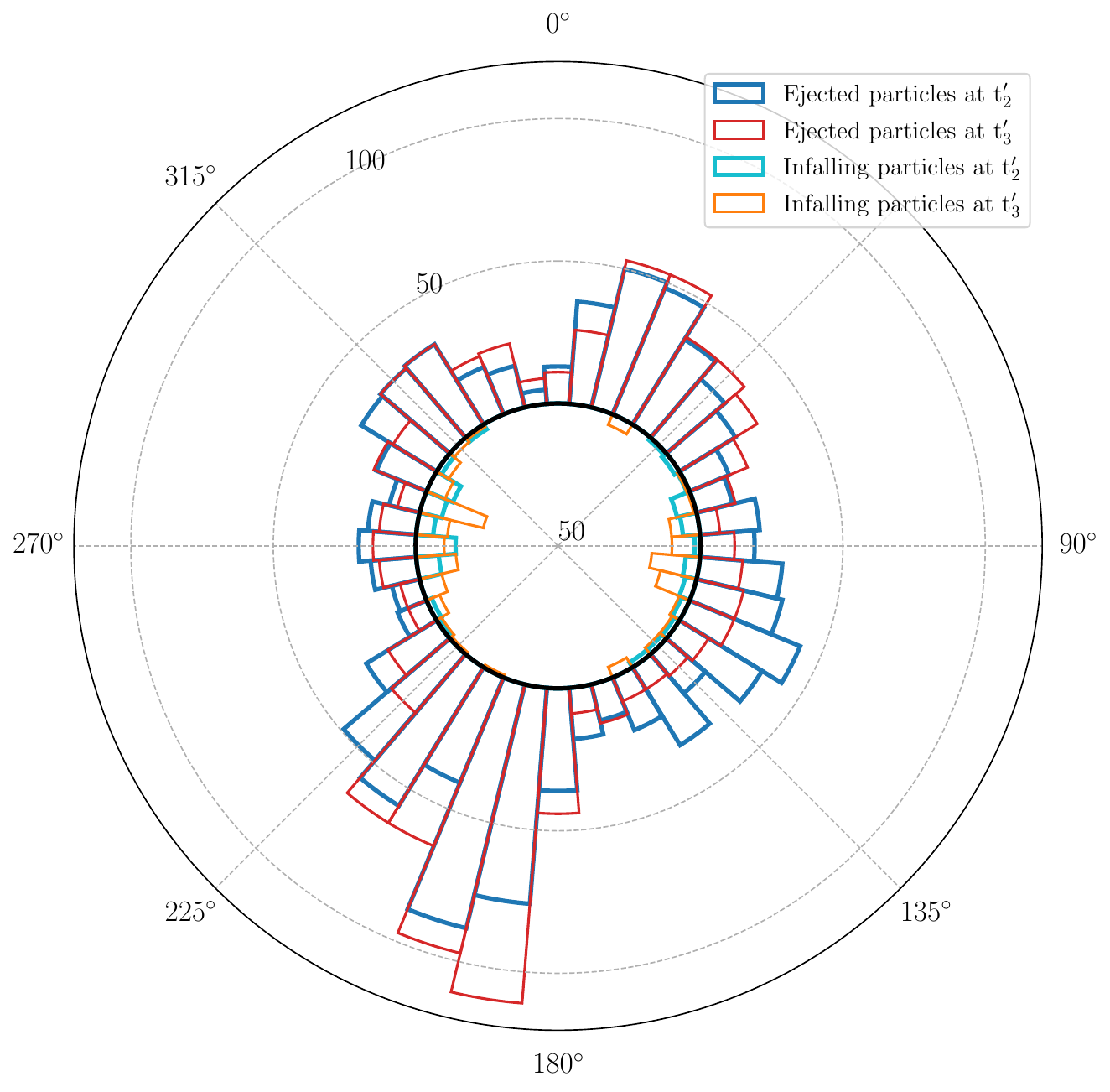}
    \end{tabular}
\caption{\textit{Top row}: Spatial distribution of the gas particles ejected by MBH feedback at $z\sim8.77$ (left-hand side) and immediately prior to the cavity opening, at $z\sim8.39$, (right-hand side). The distances of the particles from the MBH are shown in the inner circle, whereas the angular distribution histogram is reported in the outer ring. An angle of $0^\circ$ corresponds to the $z$-axis orthogonal to the galactic disc.
\textit{Bottom row}: Histograms of the ejected particle angular distribution (as number of particles) immediately after the MBH feedback event ($t_2$ and $t_2'$) and after 5~Myr ($t_3$ and $t_3'$). Outflowing particles are shown in blue and red respectively, while infalling ones are identified with cyan and orange. }
\label{fig:ang_distr}
\end{figure*}

To better understand the origin of the cavity, we analyse now the angular distribution of the gas particles ejected by the MBH in the $\sim$40~Myr preceding the formation of the cavity. We consider two specific snapshots, one at $z\sim8.77$ and one at $z\sim 8.39$. We extracted gas particles with positive radial velocities above $3000 \ \mathrm{km \ s^{-1}}$, and followed their trajectories back and forth in time. To maintain consistency over time, we align our reference systems with the galactic disc immediately prior the feedback events (with the $z$ coordinate referring to the direction perpendicular to the disc). We then flagged all the kicked particles in the time interval considered, and tracked them for about 10 Myr, at two specific times: (i) immediately after the MBH feedback event ($t_2$ and $t_2'$) and (ii) after further 5~Myr ($t_3$ and $t_3'$).

We show the spatial distribution of the identified particles in Fig.~\ref{fig:ang_distr}, before ejection in the top panels, and after ejection in the bottom ones. We find that, at both redshifts, the vast majority of the ejected gas approaches the MBH along the disc plane, but for a few particles off-plane (more at earlier times, when the galactic disc is thicker). At $z\sim 8.77$ (left panels), particles preferentially leave the system perpendicular to the disc (see the outflowing particle distribution in blue and red respectively). Nonetheless, a non-negligible number of particles is launched within the disc plane. The energy transferred to the interstellar medium in this phase becomes large enough to start regulating the MBH accretion rate, which drops to sub-Eddington rates. By $z\sim 8.39$ (right panels), the MBH-powered outflows are significantly less massive than at $z\sim 8.77$, and mostly launched in the form of radiatively driven winds with a large opening angle ($\theta =45$Â°). This makes the angular distribution of the kicked particles more isotropic, hence more prone to effectively interact with the ISM. The sequence of events just described produces a pressurised bubble that starts expanding outwards, marking the end of the rapid accretion phase of the MBH and the opening of the cavity shown in Fig.~\ref{fig:density_maps}. 
During the bubble expansion, the swept-up gas is compressed in a shell. Nonetheless, the density in this shell remains moderate, thus leaving the galaxy in a sort of `quenched phase' characterised by a moderate SFR, as seen in Fig.~\ref{fig:SFR}. The cavity expands and survives as long as the inner pressure wins over the pressure of the infalling gas. The effect of the outer pressure (and also of the reverse shock at the inner boundary of the cavity) can be observed in the bottom panels (at $t_3$ and $t_3'$, 5~Myr after the feedback has been launched), where a few of the previously kicked particles show signatures of inward radial motion. The number is  still small, however, as most particles are still found outflowing. By $z\sim 8$ (about 25~Myr after the outburst), the cavity is completely closed, and the galaxy moves back onto its original evolutionary path, removing any clear signature of `quenching'. 

\subsection{Shaping spectra: Imprints of super-Eddington accretion on galaxy observables} 

\begin{figure*}
    \centering
    \includegraphics[width=0.9\textwidth]{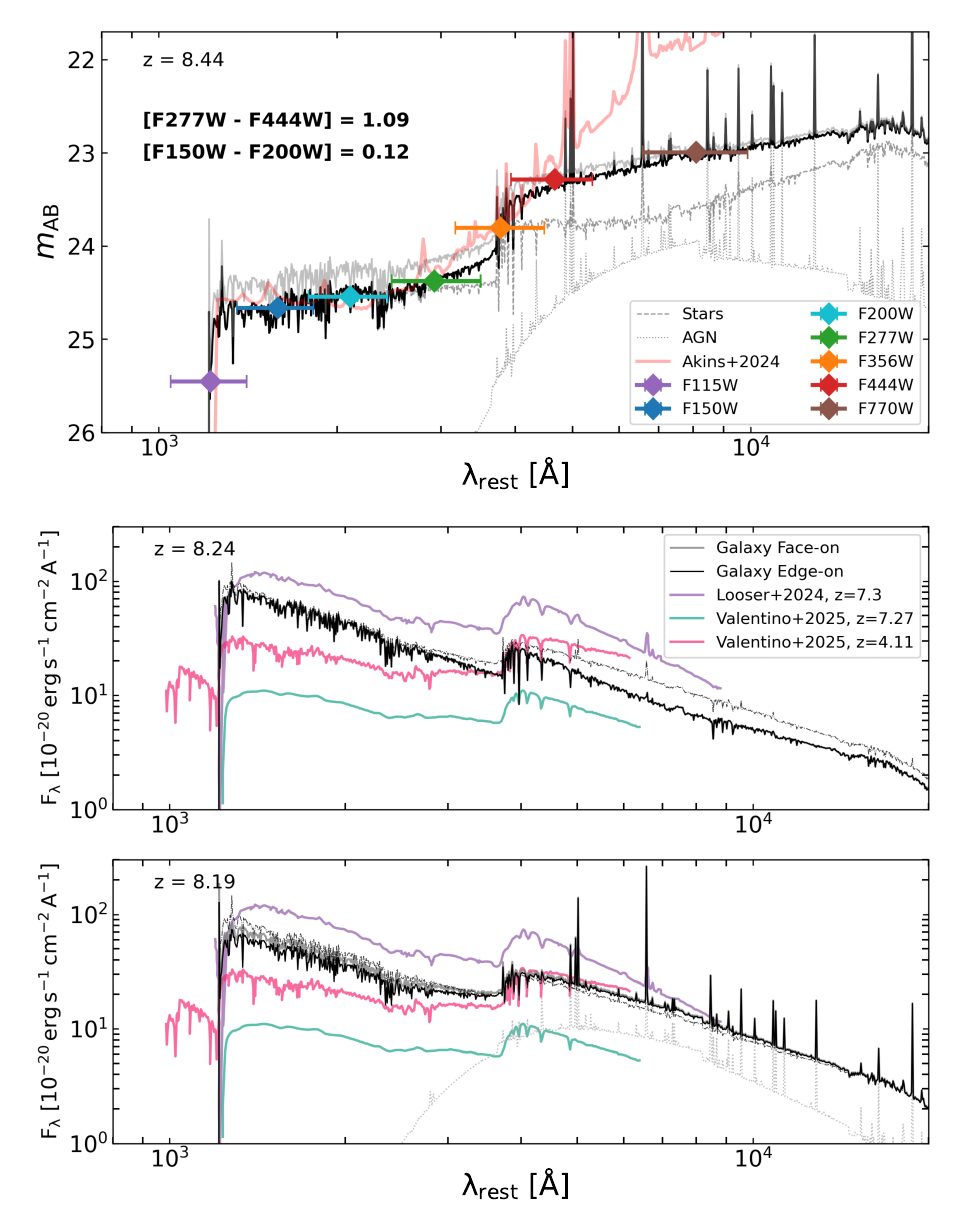} 
    \caption{\textit{Top panel}: Emerging galaxy spectrum at the end of the sustained super-Eddington growth phase as a function of the rest frame wavelength for edge-on (solid black line) and face-on (solid grey line) lines of sight. The stellar and AGN contributions (when present) are shown by dashed and dotted grey lines, respectively. Overplotted in different colours is the expected galaxy photometry in several JWST NIRCam and MIRI wide filters, which trace the rest-frame UV and optical emission. For comparison, we also show  the median-stacked SED of LRDs observed in the COSMOS-Web survey, obtained from \citet{akins24} with a solid red line.
    \textit{Central and bottom panels:} Emerging galaxy spectrum after $\approx 5 \, \rm Myr$ ($z = 8.24$) and $10 \, \rm Myr$ ($z=8.19$) from the formation of the cavity shown in Fig.~\ref{fig:density_maps}. For comparison, we include spectra of high-redshift systems recently observed undergoing phases of galaxy quenching \citep[$z=4.11$ and $7.27$]{valentino25} or mini-quenching \citep[$z=7.3$]{looser24}. For reference, in the central and bottom panels we report  the galaxy specific flux at $z=8.44$ as a thin dash-dotted black line. All spectra are rescaled to match the mass and redshift of the simulated galaxy. }
\label{fig:GalSpectra}
\end{figure*}

While the intrinsic properties of simulated galaxies give us important information about the physical modelling, they do not directly allow us to compare results with observations. For this reason, we consider here how MBH accretion and the related AGN feedback shapes the host galaxy emission, by analysing the spectral evolution and observational features associated to the simulated galaxy during and after the sustained phase of super-Eddington growth discussed in the previous Section.
We use the open-source Synthesizer package \citep{roper2025,lovell2025} \footnote{\href{https://synthesizer-project.github.io/synthesizer/}{https://synthesizer-project.github.io/synthesizer/}}, which allows to generate realistic galaxy observables and mock spectra from cosmological simulations.
The emerging spectrum of the galaxy stellar component is computed based on the stellar particle mass, age, and metallicity, including nebular emission from photoionisation reprocessing and dust attenuation in the ISM. \footnote{We assume intrinsic stellar emission from the BPASS stellar population synthesis code \citep{byrne22}, including binary stars and adopting a Chabrier initial mass function \citep{chabrier03} extending between $M_\star = 0.3 - 100\,\mathrm{M}_\odot$.}
The dust attenuation for each stellar particle is self-consistently computed along a chosen line of sight based on the gas particles distribution, SPH kernels, and metallicity, up to 1~kpc from the galaxy centre. Here we assume a Calzetti attenuation law \citep{calzetti00} and a fixed dust-to-metal ratio of 0.5 to determine the dust content of each gas particle, consistent with the assumptions within the simulation.
The AGN spectrum instead is consistently modelled starting from a multi-component accretion disc continuum \citep{mathews87}, and includes the reprocessed emission from the broad line region (BLR), narrow line region (NLR), and dusty torus, assuming dust attenuation following $A_{\lambda} \propto \lambda^{-2}$.

We also account for small-scale obscuration due the unresolved dusty torus, modelling the total optical depth as $\tau_{\rm AGN} = \tau_{\rm gal} + \tau_{\rm torus}$, where $\tau_{\rm gal}$ is self consistently measured from the galaxy gas distribution. The torus optical depth is instead estimated as
\begin{equation}
\tau_{\rm torus} = N_{\rm H, torus} \, X_{\rm H}^{-1} \, k_{\rm V} \, m_{\rm p} \, \langle Z\rangle _{\rm inn}  \, DtM  \, ,
\label{eq:NH}
\end{equation}
 where $m_{\rm p}$ is the proton mass, $X_{\rm H} = 0.76$ is the universal hydrogen mass fraction, $k_{V}$ is the effective V-band dust opacity \footnote{For simplicity, we assumed the standard value of $k_{V} = 0.0795 ~\rm M_\odot /pc^2$ used in Synthesizer to reproduce the typical Milky Way calibration between visual extinction and gas column density.}, $\langle Z \rangle_{\rm inn}$ is the average gas metallicity in the inner galaxy regions around the MBH, and $DtM = 0.5$ is the assumed dust-to-metal ratio. We evaluate the torus gas column density $N_{\rm H, torus}$ as a function of $M_{\rm BH}$ and $f_{\rm Edd}$ relying on Eq.~6 of \citet{volonteri25}, where they develop a simple, spherically symmetric model to 
 integrate the expected gas density profile from the self gravitating radius up the dust sublimation radius. We model dust emission from the galaxy as a black body at $\rm T_{\rm BB} = 50 \, \rm{K}$ \citep{tripodi23}, while assuming a warmer component with $\rm T_{\rm dust} = 100 \, \rm{K}$ for the torus, as expected if AGN obscuration occurs on small scales \citep{akins24}. However, the assumed dust temperatures have a negligible impact on our results, which will primarily focus on the UV and optical rest-frame emission of the system.
In the following, we analyse how the galaxy emission rapidly evolves on timescales of $\approx 10$ Myr due to the combined effect of AGN emission and feedback. 
Specifically, we focus on (i) the emerging galaxy spectrum right after the phase of sustained AGN accretion, and (ii) the impact of AGN feedback, in particular the formation of the wide cavity shown in Fig. \ref{fig:density_maps} and subsequent quenching, on the galaxy emission.

\subsubsection{The emergence of LRDs through super-Eddington growth}

In the upper panel of Fig. \ref{fig:GalSpectra}, we present the galaxy emerging rest-frame spectrum at redshift $z=8.44$, shortly before the formation of the gas cavity in the inner region around the central BH. Overplotted are the expected photometric measurements in different JWST NIRCam and MIRI wide filter bands, covering the rest-frame UV and optical emission. The resulting galaxy spectrum is characterised by a relatively flat UV continuum, dominated by the stellar emission, and presents at the same time a strong rising continuum in the optical rest-frame, redward of the Balmer break. This latter feature is originated by a combined contribution of older stellar populations and a moderately obscured AGN, for which we estimate an optical depth $\tau_{\rm AGN} \approx 0.98$ using Eq.~\ref{eq:NH}.

This emerging spectral shape shows a strong resemblance to the distinctive SED observed in LRDs. For a direct comparison, in Fig. \ref{fig:GalSpectra} we show the median-stacked SED of more than 400 LRDs observed between $z \sim 5 - 9$ in the COSMOS-Web survey \citep{akins24}, rescaled to align with the UV continuum of our simulated galaxy. The comparison reveals a remarkable agreement in reproducing both the flat UV spectrum and the rising optical continuum up to $\sim 500 ~\angstrom$ rest-frame. These objects have been tentatively interpreted as systems powered by mildly obscured AGNs, where super-Eddington accretion rates might help mitigate several tensions regarding their estimated MBH masses, AGN luminosities, and number densities \citep[see, e.g.][]{king24,lupi24b,lambrides24,pacucci24,madau24,rusakov25,trinca24}. However, whether their steep optical continuum is due to a combination of a stellar Balmer break and AGN emission \citep{greene24}, or results instead from a very dense, dust-free gas environment surrounding the accreting MBH \citep{inayoshi24,Ji25}, remains an open question.

In this context, it is intriguing to observe a similar spectral shape arising from the simulated galaxy after a phase of sustained super-Eddington growth. These, in particular, are expected to be accompanied by galaxy-scale outflows, such as the one presented in Sect.~\ref{sec:feedback}, which deplete the inner galactic regions of gas and clear potential lines of sight to the heavily obscured central AGN, but also to the young stellar populations. This would generate a transient phase that could explain the peculiar spectral features observed.
However, it has to be noted that the system considered here might differs significantly from the typical LRD population, being characterised by a more extended stellar distribution and hosting a central BH within the `local' scaling relation if compared to the host galaxy stellar mass \citep[$M_{\rm BH}/M_{\rm star} \sim 10^{-3}$][]{reines15}.
As a consequence, the rising spectral shape redward of $4000\, \angstrom$ is due in our case to a combination of stellar and obscured AGN emission, with the former dominating the photon budget. The resulting transition between the UV and optical emission might therefore be smoother than what could be expect in LRDs if they were powered by an overmassive black hole that significantly modifies the optical emission.
To explore this further, we compared the galaxy expected photometry in JWST NIRCam wide filters with colour selection criteria that have been employed in literature to photometrically select LRDs. These are - in the redshift range of our interest ($z \approx 8$) - $F277W-F444W > 1.0 \, \rm{mag}$ and $F150W-F200W < 0.5 \, \rm{mag}$ \citep{kocevski24}. For the simulated galaxy, we obtain values of $F277W-F444W = 1.09$ and $F150W-F200W = 0.12$, respectively, which are fully consistent with the LRD selection.

A slightly larger $M_{\rm BH}/M_{\rm star}$ ratio or a more intense accretion phase ($f_{\rm Edd} \gtrsim 1$) would likely lead to larger $F277W-F444W$ values, driving the system even closer to the typical Balmer jumps observed in the LRD population.
Naturally, the simulated galaxy is an extended system, with estimated $R_{\rm e, F444W} \approx 101 \, \rm{pc}$, which would likely not satisfy the compactness criterium considered in the search for LRDs. However, it is important to highlight that, in our analysis, we do not account for the limited surface brightness sensitivity expected in real observations. If the outer galactic regions were dim enough to fall below the JWST sensitivity, the observed galaxy half light radius would appear smaller. This would, in turn, reduce the UV contribution of the stellar component, increasing $F277W-F444W$ and steepening further the red optical continuum emission. A similar effect would occur if a fraction of the stellar population remained embedded in dense gas, even while intense AGN feedback evacuates the inner galactic regions.
Additionally, recent studies have proposed that the prominent Balmer break and absorption features commonly observed in LRDs may originate from extremely dense, dust-free, turbulent gas surrounding the massive BH \citep{inayoshi24,Ji25,naidu25}. In this scenario, the AGN emission would be intrinsically faint - rather than heavily dust-attenuated - producing a prominent Balmer jump, with the AGN dominating the emission redward of the break even for lower $M_{\rm BH}/ M_{\rm star}$ ratios. Although modelling the resulting SED in a similar scenario is beyond the scope of this work, it is worth noting that this would further enhance the consistency between the simulated galaxy and the observational selection criteria for LRDs.

The comparison between the predicted galaxy SED and the stacked spectrum of LRDs from \citet{akins24} shows that, despite the close agreement in the rest-frame UV and optical regime, the predictions start to diverge in the infrared. 
However, this tension is likely influenced by the assumption of renormalizing, for the sake of comparison, the LRD UV flux to our flat continuum. In reality, prototypical LRDs exhibit a much more moderate contribution from their host in the UV, which could explain the discrepancy in the F770W filter. Additionally, the role of strong emission lines is crucial in shaping the photometry of this observed population. In our simulated spectrum, at $z \simeq 8.4$, the strong H$\alpha$ emission falls just outside the F770W filter. Detecting the system in a comparable accretion phase at slightly higher redshift would therefore yield higher expected photometry in F770W, further alleviating the tension.
A proper comparison at longer wavelengths, however, is less reliable, as we are unable to self-consistently resolve the AGN torus emission in the simulation, and therefore we have to assume a fixed dust temperature. Nevertheless, this increasing tension in the mid-infrared highlights the importance of dedicated observations, and especially deep MIRI spectroscopy, to probe the nature of the steep red optical continuum. Recent studies have started to place more stringent constraints on the dust content of these systems, suggesting that it may be difficult to reconcile with the heavily obscured AGN scenario and may instead favour dust-free, gas-enshrouded MBHs \citep{perez-gonzalez2024,chen2025,setton2025}.

However, in our simulation, as shown in Fig \ref{fig:SFR}, the prolonged phase of super-Eddington accretion, lasting $\approx 50 \rm \, Myr$, triggers a temporary quenching of the host galaxy, matched by a sharp decline in the MBH accretion rate. This suggests that the described spectral feature represents a transient phase in early galaxy evolution, linked to brief episodes of accelerated MBH growth. Notably, this phase may also precede a short-lived quenching of the host galaxy, as discussed in the following section.

\subsubsection{Are JWST galaxies quenched or just playing dead?}

As extensively discussed in Sect.~\ref{sec:feedback}, after the prolonged phase of efficient accretion onto the simulated MBH, a $\gtrsim 200 \, \rm pc$ cavity is opened in the gas distribution. This brings the galaxy into a state of mild quenching that lasts for $\lesssim 100 \, \rm Myr$ before recovering the previous sustained SF activity.
To investigate how this rapid transition affects the galaxy emission, and whether the temporary quenching leads to characteristic observational features, we show the galaxy spectrum at $z=8.24$ and $z=8.19$ in the bottom panels of Fig. \ref{fig:GalSpectra}, 
corresponding to $5$ and $10 \rm \, Myr$ after the formation of the central gas cavity.
The first snapshot captures the peak of the impact of the AGN feedback event. In this phase, the SFR has decreased by a factor of $\sim 3$ from previous values of $\sim 200 \, \rm M_\odot/yr$, while the AGN activity is starving, as additional gas inflows still haven't refuelled the nuclear regions.
In the second snapshot, the SFR remains at comparable low levels, while the central AGN starts to reactivate.

For comparison, in Fig. \ref{fig:GalSpectra} we include the spectra of several high-redshift systems recently observed with JWST, which have been identified as undergoing a phase of galaxy quenching.
We first compare our simulated galaxy to the systems presented in \citet{looser24}, a $z = 7.3$ galaxy which appears to have recently experienced a `mini-quenching' episode on a $ \lesssim 50 \, \rm Myr$ timescale, following a $\approx 100 \, \rm Myr$ phase of efficient SF. While both SN and AGN feedback might be effective in quenching a galaxy of this mass ($M_{*} \sim 10^{8.3} \, \rm M_\odot$), the detection of a relatively low average metallicity ($Z/Z_\odot \sim -2$) tentatively supports the AGN-feedback scenario. These mini-quenching phases, likely triggered by fast and powerful outflows rather than slower quenching mechanism, are expected to be short-lived. As a result, they represent only a transient phase in the galaxy evolution and are therefore exceptionally rare to observe, especially in the high-redshift Universe.
This process is closely reproduced in our simulation, where we predict a temporary drop in SFR lasting $\sim 50-100 \, \rm Myr$ after a phase of intense super-Eddington accretion onto the central MBH.
Focusing on the spectral comparison, we observe a remarkably consistent spectral shape between the observed galaxy and the simulated system at $z=8.24$, $\sim 5 \, \rm Myr$ after the formation of the cavity. Both present a blue UV slope and a relatively mild Balmer break, which would still photometrically classify them as `star-forming' systems, despite their recent quenching.

We also include the spectra of the systems analysed in \citet{valentino25}, where outflows are detected in two massive galaxies ($M_{*} \sim 10^{10.2} \, \rm M_\odot$) at $z= 4.11$ and $z = 7.27$. These galaxies show signatures of recently quenched star formation, such as the lack of strong emission lines and reconstructed $\rm SFR_{\rm 100 \rm Myr}\lesssim 15 \, \rm M_\odot/yr$. Interestingly, neither system shows clear evidence of ongoing AGN activity, despite the high mass-loading factor ($\eta \approx 50$) estimated for the higher-redshift one suggests AGN feedback as powering mechanism for the observed outflow.
When comparing their spectra to our simulated galaxy, we find a smoother UV slope in the observed systems, which likely reflects a more extended phase of quenching.
However, a notable evolution is observed in the simulated spectra shortly after, roughly $\sim 10 \rm \, Myr$ after the formation of the cavity. At this point the galaxy SFR remains low, while the central AGN starts to re-activate thanks to new gas inflow on nuclear scales. In this phase, the - subdominant - AGN contribution boosts the optical rest-frame continuum, leading to redder photometric colours. In the lower panel of Fig. \ref{fig:GalSpectra} we observe how this effect improves the alignment between the simulated spectra and the ones provided in \citet{valentino25}, suggesting that it could potentially mimic a longer duration of the galaxy quenching phase retrieved relying on SED fitting techniques. Note, however, that this contribution might be accompanied by typical features of a reactivated AGN, such as high line ratios or broad emission lines, which we do not consider in our comparison.

It is interesting to note that the in-depth analysis of a prototypical LRD recently presented in \citealt{naidu25} suggests that the underlying galaxy emission, once disentangled from that of the central AGN, closely resembles that of a mini-quenched system. This result offers the first tentative indication of a possible link between a peculiar growth phase characterizing the central MBH and its influence on the host galaxy properties.

We emphasise once again that, as in the previous section, we are not claiming that the simulated galaxy precisely reproduces the systems observed by JWST. Rather, this comparison suggests that similar episodes of galaxy quenching might represent short-lived transient phases in the early galaxy evolution, potentially triggered especially by phases of super-Eddington growth of the nuclear MBH, which are characterised by the emergence of strong winds and/or collimated jets.
Interestingly, our simulation suggests that in galaxies with $M_{\star} \sim 10^{10} \, \rm M_\odot$, even maintaining elevated SFR of $\sim 200 \, \rm M_\odot/yr$ for $\gtrsim 100 \, \rm Myr$ - which are one order of magnitude above the averaged estimated values for the observed galaxies considered here - the stellar feedback is not sufficient to trigger extended phases of quenching, while being the main responsible for regulating the SF. While our system evolves at earlier cosmic epochs and is embedded in a high density peak, which favour strong gas inflows and accelerated mass assembly potentially damping the integrated effect of SN feedback, this also suggests that similar episodes of temporary galaxy quenching might more likely be linked to AGN activity.

\section{Conclusions}
In this study, we have analysed the zoom-in cosmological simulation presented in \citet{lupi24}, with the aim of assessing the impact of super-Eddington accretion on the (co-)evolution of the galaxy host and the central MBH. We can summarise our results as follows:

\begin{itemize}
    \item Despite the stronger MBH feedback compared to L19, especially during super-Eddington phases when the total energy conversion efficiency is about 50-60 percent, the MBH does not significantly affect the galaxy host. This is due to the fact that the ejected material preferentially leaves the system perpendicular to the galactic disc and interacts moderately with the galaxy. The only exception is a relatively short ($\sim 50$~Myr) quenching event during which the SFR drops by up to one order of magnitude, which slows down the stellar mass build-up.
    \item As soon as the MBH exceeds $M_{\rm BH}\sim 10^6\ \msun$, the impact of the accretion-powered feedback, both in the form of radiation and kinetic winds or jets, starts to dominate over SN feedback, as shown in Fig.~\ref{fig:luminosity}. However, for the subsequent 70~Myr (down to $z\sim 7.8$) the  SFR does not significantly change (see Fig.~\ref{fig:SFR}). This, together with the sustained accretion rate of the MBH, suggests that AGN feedback in this phase only affects a limited region of the galaxy, of the size of the accretion sphere of the MBH itself (a few tens of parsec on average), preferentially launching outflows that do not significantly alter the galaxy interstellar medium. We therefore conclude that SF across the galaxy is mostly regulated by stellar feedback. Around the end of the simulation, the MBH accretes close to the Eddington limit and launches outflows that are preferentially  perpendicular to the galactic disc (see last panel of Fig.~\ref{fig:ang_distr}), which again results in a limited impact on the galaxy-wide SF compared to stellar feedback. 
    \item The huge kinetic power of winds and jets from the MBH results in the ejection of significant gas from the galaxy centre but only a fraction of it is fast enough to leave the halo, with the rest producing galactic fountains. This is particularly evident during the outburst phase around $z\sim 8.3$, when the continuous pushing of the AGN winds and jets on to the galactic disc resulted in the creation of a cavity a few hundred parsecs  in size, which survived for about 50~Myr. This event that almost shut-off SF also stopped MBH accretion, until new gas started flowing towards the centre, rebuilding the galactic disc, and starting a new accretion phase for the MBH at roughly the Eddington rate;
    \item With our model, we were able to directly probe the multi-phase nature of accretion-driven outflows, which are dominated by slower molecular and neutral gas within the central few kiloparsecs, whereas faster ionised outflows start to dominate at larger distances. A comparison with observed outflows in high-redshift systems showed that our estimated outflow rates and velocities are in line with expectations when suitably rescaled to the simulated MBH mass and luminosity.
    \item Relying on the \texttt{Synthesizer} package to post-process the galaxy emission, we found that in the evolutionary phase following MBH super-Eddington growth, the quasar host progenitor exhibits the typical spectral shape observed in LRDs. This is likely the result of a combined effect of obscured AGN emission and mounting AGN feedback, which clears lines of sights towards the inner galactic regions. The agreement with standard LRD colour selection criteria is marginal, due to (i) the relatively low MBH-to-stellar mass ratio of the simulated system compared to what is expected for this emerging population of sources, (ii) the larger physical size of our simulated galaxy ($R_{\rm h} \sim 100 \rm \, pc$). However, when considering realistic sensitivity limits of JWST observations, the galaxy outskirts might be too faint to be detected, which would slightly reduce the stellar contribution and bring the system into closer agreement with the expected selection criteria. The same would hold if the system were observed during phases of more sustained accretion, with $f_{\rm Edd} \gg 1$.    \item During the major outburst event observed in the simulation, which follows a prolonged phase of super-Eddington growth, the galaxy undergoes a rapid transition ($\sim 5-10 \, \rm Myr$) in its spectral emission due to gas depletion in the central galactic regions. In this phase, the post-processed emission closely reproduces the spectra observed in high-redshift systems that undergo phases of quenching \citep{valentino25} or mini-quenching \citep{looser24} at $z\approx 4 - 7$. This similarity shows how these events might just represent transient phases during galaxy evolution,  which are expected to be relatively short-lived ($\sim 1/10^{\rm th}$ of the galaxy lifetime at $z\sim8$) especially at high-redshift, where abundant gas supply and strong inflows can rapidly rejuvenate star formation. The simulation shows that, for massive galaxies, these temporary quenching episodes are more likely driven by AGN feedback events, possibly related to a period of enhanced MBH growth.
    
\end{itemize}
In conclusion, we have demonstrated that our simulations can capture the evolution of high-redshift quasar host progenitors and the interplay with the central MBH in detail. In particular, we have been able to show that these systems, despite having been studied for more than a decade, remain poorly understood, especially in their infancy phases, when they can change their appearance over extremely short timescales and mimic diverse classes of objects that JWST has revealed for the first time. These rapid transitions underscore the highly dynamic nature of quasar host progenitors (and massive galaxies in general) in the early Universe and highlight the need for particular caution when interpreting the observational properties of high-redshift sources and inferring their long-term evolution.

\begin{acknowledgements}
We thank the referee for constructive comments and suggestions that helped us to improve the manuscript. GQ acknowledges support by the FARE2020 `CosmicNodes'(R20S99FS3J) project. AT and AL acknowledge support by the `PRIN MUR 2022935STW' funded by European Union-Next Generation EU, Missione 4 Componente 2, CUP C53D23000950006. AT acknowledge financial support from the Bando Ricerca Fondamentale INAF 2023, Mini-grant `Cosmic Archaeology with the first black hole seeds'.
\end{acknowledgements}
\bibliographystyle{aa}
\bibliography{./Biblio}

\begin{appendix}

\section{Details on the numerical implementations}
\label{app:methods}
\subsection{Gas thermodynamics and chemistry}
As in Paper I \citep{lupi19} and II \citep{lupi22}, chemical abundances are evolved out of equilibrium together with heating/cooling via the publicly available astrochemistry package \textsc{krome} \citep{grassi14}. Precisely, the non-equilibrium chemistry of a more detailed chemical network of $29$ distinct species has been treated, including H, He, H$_{2}$, O, Si, C, Fe, N, and most of their ionised states (C[I-IV], O[I-VI], N[I-V], and Fe[I-II]), with some of them commonly observed in quasar hosts. As for heating/cooling processes, we employ the same processes discussed in \citet{lupi22}. To accurately model cooling at low temperatures in low-metallicity environments and to properly describe the different phases of the interstellar medium (ISM), the simulation also includes H$_2$ formation and evolution, specifically in the gas phase via H$^-$ detachment and on dust grains, modelled assuming a fixed dust-to-gas ratio linearly scaling with the gas metallicity (see \citealt{lupi18} for a detailed description). 
Due to the limited resolution, the clumpy structure of molecular clouds where stars are expected to form cannot be resolved accurately, thus resulting in an underestimated formation rate of H$_2$ \citep{lupi18}. For this reason, a sub-resolution clumping factor $\mathcal{C}_{\rho}=\textrm{exp}({\sigma_{s}^{2}})=1 + b^2 \mathcal{M}^2$ is included in the formation rate of molecular hydrogen on dust in the simulated high-density regions. Here, $\mathcal{M}$ is the Mach number and $b$ is a parameter associated to the turbulence modes in the cloud, set to $0.4$, which corresponds to a statistical mixture of compressive and solenoidal modes. H$_{2}$ dissociation is also included, accounting for both the direct branch and the Solomon process.

Photochemical reactions and photoheating are also included, in the form of a suitably shielded uniform metagalactic UV/X-ray diffuse background \citet{haardt12}, and a local radiation from young massive stars added on top. As a further improvement, in order to take into account the hot X-ray corona that commonly surrounds MBHs, X-ray photochemistry has also been included, as extensively discussed in Appendix \ref{app:xrays}.
Finally, in order to simplify the calculations at high temperatures (where abundances quickly reach chemical equilibrium), for metal cooling above $T=10^4$~K we employ pre-computed look-up tables \citep{shen13} obtained with the photoionisation code \textsc{CLOUDY} \citep{ferland13} under the assumption of photoionisation equilibrium with the UV background.

\subsection{Star formation}
Due to the limited mass resolution, stellar particles in the simulation represent entire stellar populations distributed according to a Kroupa initial mass function \citep{kroupa2001}. Star formation (SF) is then implemented using a stochastic prescription, which converts gas into stellar particles according to a star formation rate (SFR) density
\begin{equation}
\dot{\rho}_{\rm SF}=\epsilon \frac{\rho_{\rm g}}{t_{\rm ff}}
\end{equation}
where $t_{\rm ff}=\sqrt{3\pi/(32G\rho_{\rm g})}$ is the free-fall time, $\rho_{\rm g}$ the local density of gas, and $G$ the gravitational constant. In this specific case, the SF efficiency parameter $\epsilon$ is allowed to vary during the evolution, according to the results of turbulent magnetised clouds by \citet{padoan12}, as in \citet{lupi20}. Despite the model not requiring any density threshold, we allow star formation (SF) only for gas above $1 \, \mathrm{cm^{-3}}$ to avoid the calculation of the SF efficiency in regions where SF will never occur. When the SF criteria are met, a stellar particle is stochastically spawned with a mass equal to that of the progenitor gas cell.

\subsection{Stellar feedback}
Stellar feedback can be either radiative or mechanical. Regarding the latter, both type II and type Ia supernovae (SNe) are considered in the simulations. We assume that, after $\sim 5$~Myr, the most massive stars, i.e. $8 \msun<M_{\star}<40 \msun$,  explode as type II SNe, while type Ia SNe, that are associated to stellar binary systems, explode according to a delay time distribution \citep{maoz12}. Their explosions directly affect the gas content of the halo, together with the metallicity of the gas, and are modelled as discrete events based on the prescription presented in \citet{lupi20} and improved as discussed in \citet{lupi24}. In addition to SNe, we also include the mass, metal, and energy release by stellar winds (from massive stars before SN explosions and from stars below $8 \msun$; see \citealt{lupi24} for details).

In addition, we also include radiative feedback from stars, in particular from young massive stars producing strong ionising and dissociating UV flux able to affect the atomic and molecular gas thermodynamics. Instead of the cost-effective approximated radiation transport of L19, we include here on-the-fly radiation transport as largely described in \citet{lupi20b}, with the reduced speed of light set to $c_{\rm red}=1000\rm\, km\, s^{-1}$, which is large enough compared to the gas motions in order not to affect our results. Photons are coupled to the gas in the thermochemistry step, where photochemical reactions and heating are accounted for through the package \textsc{krome}, as discussed in \cite{lupi18}. As in this simulation the radiation spectrum also extends to X-rays, we track radiation in 11 photobins ranging from 0.7 eV up to 10 keV, where two bins are used to cover soft (0.2-2 keV) and hard (2-10 keV) X-rays.\footnote{The radiation spectrum is sampled with 11 bins with energy limits  [0.7, 1.6, 5.6, 11.2, 13.6, 15.2, 24.59, 54.42, 100.0, 200.0, 2000.0, 10000.0]. }

\subsection{MBH physics}
Here, we summarise the prescriptions we implemented in \gizmo \ to model MBH seeding, growth and feedback, and MBH binary mergers (see \citealt{lupi24} for details).

\subsubsection{Seeding}
Since the resolution in our simulation is not high enough to track different formation mechanisms of the MBHs, we employ here a simple `seeding' prescription, in which a MBH seed with an initial mass $M_{\rm BH}=10^5\msun$ is created in those galaxies having at least $10^8\ \msun$ in stars, identified via a Friends-of-Friends algorithm (see L19). As is common in most cosmological simulations, the dynamical friction that drives a MBH to the centre of its host galaxy is not always resolved during the evolution of the system, especially in the early stages when the mass and spatial resolution are insufficient. This can potentially result in spurious MBH scattering. As described in \citet{lupi24}, we include here an artificial correction for the unresolved dynamical friction effect \citep{dubois13,tremmel15,pfister19}. In order to guarantee a large enough mass ratio between the MBH and the other particles \citep{tremmel15}, which ensures a more accurate dynamical evolution, we also decouple the MBH mass in a physical mass (used for accretion, see above) and a dynamical mass (employed for the dynamics), the latter set to a value ten times larger \citep{anglesalcazar17}. The gap between the two masses can be physically justified as an unresolved stellar envelope surrounding the MBH, commonly found in many galaxies in the local Universe \citep[see][for a review]{neumayer20}.

\subsubsection{Accretion} \label{BHL}
Because of the limited resolution in our simulations, which does not guarantee the influence radius of MBHs to be always resolved, we employ the common Bondi-Hoyle-Lyttleton (BHL) accretion formula \citep{hoyle39, bondi44, bondi52}:
\begin{equation}
\dot{M}_{\rm BH}=\dot{M}_{\rm BHL}\equiv\frac{4 \pi G^2 {M_{\rm BH}}^2 \rho_{\rm gas}}{\left( {v_{\rm rel}}^{2} + {c_{s}}^{2}\right)^{3/2}}
\end{equation}
where $M_{\rm BH}$ is the MBH mass, $\rho_{\rm gas}$ is the density of the gas surrounding the MBH, $v_{\rm rel}$ is their relative velocity, and $c_{s}^{2}$ is the sound speed. In order to prevent the MBH from accreting low-density material located far away from it, the MBH accretion length, defined by the radius encompassing 96 neighbours, is capped at a maximum physical radius $h_{\rm max}=1$ kpc.
As detailed in \citet{lupi24}, we do not cap the accretion rate at the Eddington limit.

\subsubsection{Feedback} 
\label{sec:feedback}
MBH feedback is implemented in both radiative and kinetic forms, with the feedback mode adjusted consistently based on the accretion regime.
We distinguish the different regimes in terms of the Eddington ratio $f_{\rm Edd}=\dot{M}_{\rm BH}/\dot{M}_{\rm Edd}$, where $\dot{M}_{\rm Edd}=16\, L_{\rm Edd}/c^2$ and $L_{\rm Edd}=1.38\times 10^{38}(M_{\rm BH}/\msun)\rm\, erg\ s^{-1}$ is the Eddington luminosity \citep{madau14}.
As described in detail in \cite{lupi24}, radiative feedback is injected assuming a composite blackbody + X-ray corona spectrum, implemented using an approximate fit that depends on the MBH mass and bolometric luminosity. The radiative efficiency is defined as a function of both the MBH spin and the Eddington ratio \citep{madau14,lupi16}, and is computed in the simulation assuming a constant spin $a=0.7,$ always aligned with the angular momentum of the gas surrounding the MBH \citep[see, e.g.][for a similar implementation where spin is also evolved]{husko25}. 
Kinetic feedback is instead implemented by imparting a kick with $v_{\rm kick}=30000$~km/s to gas particles within the MBH kernel. The feedback efficiencies (radiative and kinetic) are defined for the different accretion regimes as follow \citep[see][for details]{lupi24}:
\begin{itemize}
    \item in the advection dominated accretion flow regime \citep[$f_{\rm Edd}<2.5 \times 10^{-3}$]{yuan14}, the radiative efficiency is set to $\eta_{\rm rad} = 0.103 \ f_{\rm Edd}^{2/3}$ \citep{xie12}. A jet is launched along the MBH spin direction with an efficiency $\eta_{\rm jet}(a, \Phi) \simeq 0.417 \ \Phi^2\ $, where $\Phi\equiv \phi/\phi_{\rm MAD}$, $\phi$ is the magnetic flux in the disc, and $\phi_{\rm MAD}$ is the magnetic flux in a magnetically arrested disc \citep[MAD,][]{narayan03}. 
    \item for $2.5 \times 10^{-3}<f_{\rm Edd}<1$, we assume a standard \citep{shakura73} disc solution, with $\eta_{\rm rad}=0.103$, which corresponds to the value of the disc radiative efficiency $\eta_{\rm disc}\equiv 1-\sqrt{1-2R_{\rm ISCO}(a)/3}$ for our assumed spin value $a=0.7$. In addition, we include a radiatively driven wind (produced on unresolved scales). Since cooling is expected to be efficient \citep{king03}, we assume the wind to be in the momentum conserving regime, which translates in a kinetic efficiency $\eta_{\rm kin}=0.05$. Unlike jets, AGN winds are expected to subtend a large solid angle, that we set at 45Â° of semi-aperture with respect to the MBH spin direction.
    \item in the super-Eddington regime ($f_{\rm Edd}>1$), we assume the slim-disc solution with $\eta_{\rm rad}$ determined as in \citet{lupi16}, and the production of a jet with $\eta_{\rm jet}(a, \Phi) \simeq 0.637 \ \Phi^2$.
\end{itemize}

\subsection{Initial conditions}
The initial conditions of the simulations are the same discussed in L19, and have been generated via MUSIC \citep{hahn13} at $z=100$, assuming a box size $75 \ \rm Mpc \ h^{-1}$ and a course resolution grid of $256^{3}$.
The target halo was accurately chosen to match the typical halo mass \citep[$\sim 3\times 10^{12}\rm\,\msun$;][]{dimatteo17, tenneti18} at $z=6$ and the galaxy overdensity \citep{uchiyama18, mignoli20} expected for high-redshift quasar hosts. Since only one halo with this mass is expected to form on average in our cosmological box, we opted to employ a `constrained realisation' technique to ensure at least one was found. For our simulations, we adopted the \citet{planck16} cosmological parameters $H_{0}=67.74 \ \rm km \ s^{-1} Mpc^{-1}$, $\Omega_{m}=0.3089$, $\Omega_{\Lambda}=0.6911$, $\Omega_{b}=0.0489$, $\sigma_{8}=0.8159$, $n_{\rm s}=0.9667$, assuming negligible contribution from both radiation and curvature. 
From the parent dark-matter-only simulation, we recursively zoomed-in on a Lagrangian volume extending up to 2.5 virial radii of the target halo, following the approach by \citet{fiacconi17}, which resulted in  an equivalent resolution of $8092^3$ in the high-resolution.

\section{X-ray chemistry}
\label{app:xrays}
Unlike UV photons, X-rays can penetrate deeper into dense gas, because of their small cross-section. As a consequence, their net contribution to the primary ionisation process is small. However, because of their high energy, the energy of the primary electrons produced by ionisation is enough for the electrons to collide with nuclei and further ionise the gas in the so-called `secondary ionisations'. To date, two effective models have been proposed to describe secondary ionisations, and the related Coulomb interaction heating, one by \citet[][S85 hereafter]{shull85} for a mixture of H and He, and one by \citet[][D99 hereafter]{dalgarno99} for H, H$_2$, and He. Since in our chemical network H$_2$ is particularly relevant, D99 is better suited for our purpose. However, the modelling by D99 is only valid for ionisation fractions up to $x_{\rm e}\sim 0.1$, whereas the formalism by S85 can be easily extended to higher $x_{\rm e}$. In the following, we reformulate the results by D99 for a mixture of H, He, and H$_2$ using the formalism in S85, providing new fits that can be easily integrated in numerical calculations.
\subsection{Secondary ionisations}
The photon interaction with chemical species can be generally described by a photoionisation rate and a heating rate, defined as 
\begin{eqnarray}
    \zeta_i &=& \int_{E_{\rm min}}^{E_{\rm max}} \sigma_i(E)\frac{J(E)}{E}dE \equiv \int_{E_{\rm min}}^{E_{\rm max}} \zeta_i(E)dE\\
    H_i &=& \int_{E_{\rm min}}^{E_{\rm max}} \sigma_i(E) \frac{J(E)}{E}(E-E_{\rm th})d E,
\end{eqnarray}
where $E_{\rm min}$ and $E_{\rm max}$ are the limits of the photon energy band considered, $\sigma_i(E)$ is the $i$-th species photoionisation cross section at a photon energy $E$, and $\zeta_i(E) = \sigma_i(E)J(E)/E$.
Now, the secondary ionisation rate can be expressed as (S85)
\begin{equation}
    \zeta_{i,\rm sec}(E) = \frac{1}{n_i}\sum_j \phi_i(E)   n_j\zeta_{j}(E)
\end{equation}
where $\phi_i(E)$ is the photon-energy dependent number of secondary ionisations per primary electron, and $n_i$ and $n_j$ are the $i$-th and $j$-th species number densities. If we integrate over the X-ray spectrum and introduce the $i$-species fractional abundance $x_i=n_i/n_{\rm H_{tot}}$, we get
\begin{equation}
    \zeta_{i,\rm sec} = \frac{1}{x_i} \sum_j \int_{E_{\rm min}}^{E_{\rm max}} \phi_i(E)\frac{J(E)}{E}  \frac{n_j}{n_{H_{tot}}} \sigma_j(E) d E
\end{equation}
Now, we can invert the order of the operations, obtaining
\begin{equation}
\begin{split}
    \zeta_{i,\rm sec} &= \frac{1}{x_i} \int_{E_{\rm min}}^{E_{\rm max}} \left( \sum \frac{n_j}{n_{H_{tot}}} \sigma_j(E) \right) \phi_i(E)\frac{J(E)}{E}   d E \\ 
    &= \frac{1}{x_i} \int_{E_{\rm min}}^{E_{\rm max}} \bar{\sigma}(E) \phi_i(E)\frac{J(E)}{E} dE,
\end{split}
\end{equation}
where we replaced $\sum n_j/n_{H_{tot}} \sigma_j(E) \equiv \bar{\sigma}$. This result is analogous to the expressions in D99, when one considers $W_i(E) = \frac{E}{\phi_i(E)}$ \citep[see][for details]{dalgarno99}.
Now, instead of using the photon-energy dependent values, we consider the flux-averaged value $\langle\phi_i\rangle$ (or the equivalent $\langle E/W_i(E)\rangle$), as in \citet{meijerink05} and \citet{meijerink12}, and approximate the rate as
\begin{equation}
    \zeta_{i,\rm sec}' \simeq \frac{\langle\phi_i\rangle}{x_i}  \int_{E_{\rm min}}^{E_{\rm max}} \bar{\sigma}(E)\frac{J(E)}{E}d E = \frac{\langle\phi_i\rangle}{x_i} \bar{\zeta},
\end{equation}
where $\bar{\zeta}=\sum_j x_j\zeta_j$.

Now, we need to determine the values of $\langle \phi_i\rangle$ for our mixture of gas. To this aim, we assumed the limiting value of $W_i(E)$ for very-high energies \citep{meijerink05,meijerink12}, thus considering the highest energy available in D99 (1~keV), and re-fitted the data in D99 using the functional form by S85, obtaining
\begin{eqnarray}
    \phi_{\rm H\,\,}(E) &=& \phi_{\rm H,0}(E) / (1+1.89n_{\rm H_2}/n_{\rm H})\\
    \phi_{\rm H_2}(E) &=& \phi_{\rm H_2,0}(E) / (1+0.53n_{\rm H}/n_{\rm H_2})\\ 
    \phi_{\rm He}(E) &=& \phi_{\rm He,0}(E)
\end{eqnarray}
where 
\begin{eqnarray}
    \phi_{\rm H,0\,\,}(E) &=& \left[\frac{E}{13.6\rm\, eV}-1\right]0.3443(1 - x_{\rm e}^{0.6649})^{4.3928}\\ 
    \phi_{\rm H_2,0}(E) &=& \left[\frac{E}{15.4\rm\, eV}-1\right]0.3749(1 - x_{\rm e,eff}^{0.5805})^{2.3414}\\
    \phi_{\rm He,0}(E) &=& \left[\frac{E}{24.6\rm\, eV}-1\right]0.0509(1-x_{\rm e}^{0.7883})^{4.9301}\\
\end{eqnarray}
and $x_{\rm e,eff} = 1.83x_{\rm e}/(1+0.83x_{\rm e})$.

\begin{figure}
    \centering
    \includegraphics[width=\columnwidth]{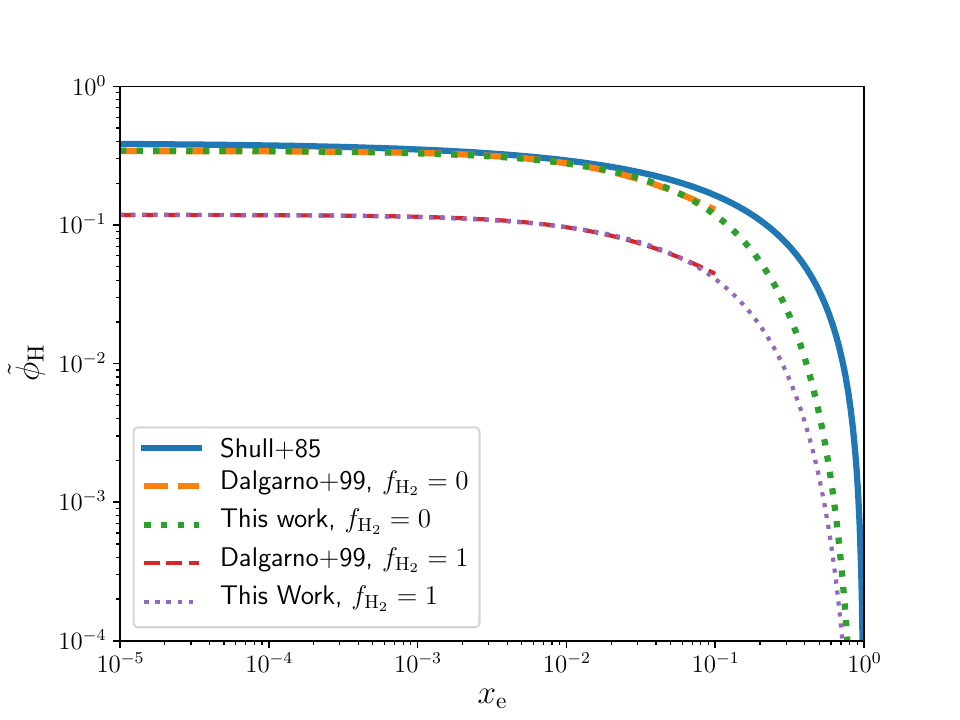}
    \caption{Number of secondary ionisations for primary electron of H. We compare the results by S85 (blue solid lines) with those by D99 (orange thick and red thin lines) for $f_{\rm H_2}=0$ and 1 respectively, and with our new fits (green thick and purple thin dotted lines).}
    \label{fig:fitsH}
\end{figure}

\begin{figure}
    \centering
    \includegraphics[width=\columnwidth]{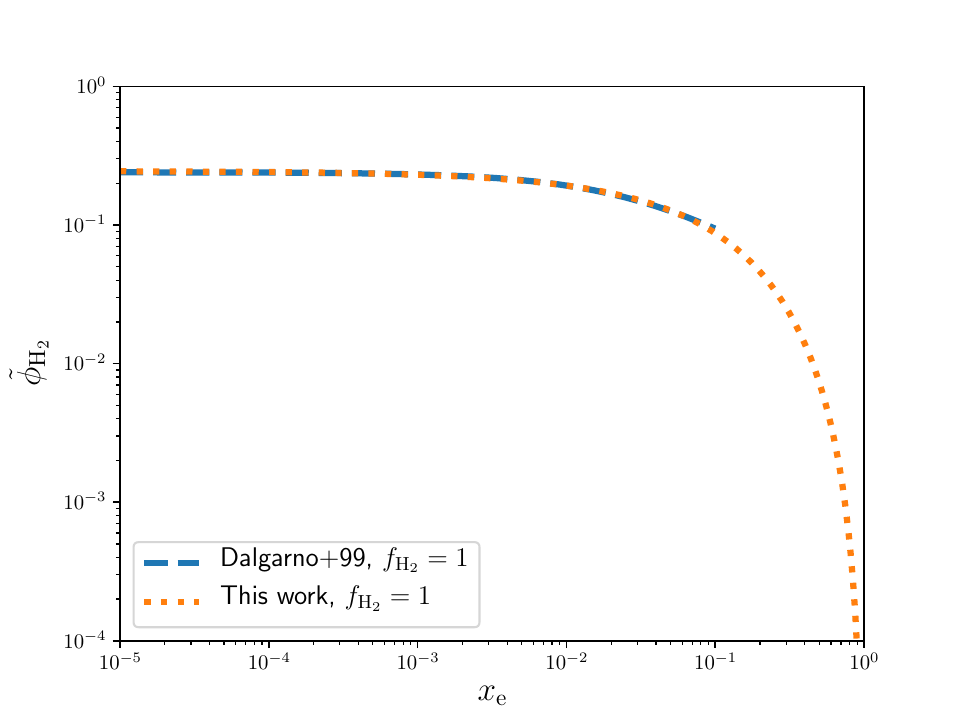}
    \caption{Same as Fig.~\ref{fig:fitsH}, for H$_2$.}
    \label{fig:fitsH2}
\end{figure}

\begin{figure}
    \centering
    \includegraphics[width=\columnwidth]{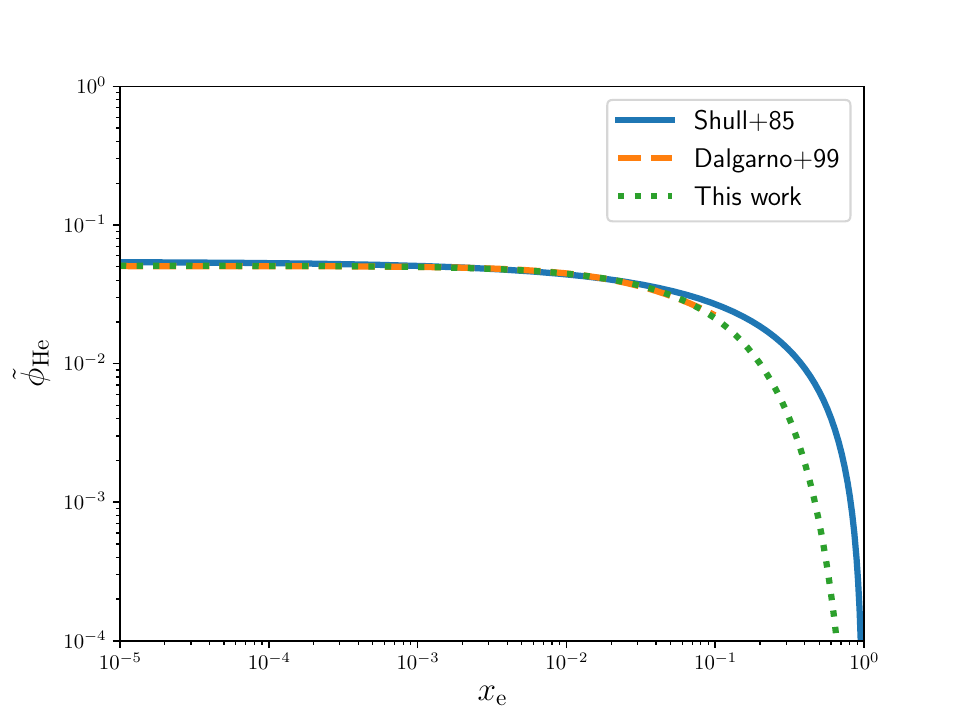}
    \caption{Same as Fig.~\ref{fig:fitsH}, for He.}
    \label{fig:fitsHe}
\end{figure}

In Fig.s~\ref{fig:fitsH},~\ref{fig:fitsH2}, and~\ref{fig:fitsHe}, we compare our new fits to $\phi_i$ with the original works by D99 (shown as orange thick and red thin lines) and S85 (shown as blue solid thick lines). The top panel reports $\tilde{\phi}_{\rm H} = \phi_{\rm H}/(E/13.6-1)$ for $f_{\rm H_2} \equiv n_{\rm H_2}/n_{\rm H}  = 0$ (thick lines) and $f_{\rm H_2}=1$ (thin lines), the middle one shows $\tilde{\phi}_{\rm H_2}$, and the bottom one $\tilde{\phi}_{\rm He}$. Our new fits are reported as green thick and purple thin dotted lines. In the interval where both S85 and D99 are present, our new fits agree very well with D99, with mild differences from S85, whereas for higher $x_{\rm e}$, they naturally extend smoothly up to the fully ionised case.  

For heavy elements, secondary ionisations from primary electrons can also be important. To include them, we follow \citet{meijerink05}, \citet{meijerink12}, and \citet{adamkovics11}, rescaling the hydrogen rate $\zeta_H$ by the cross-section ratio $\sigma_{\rm A}/\sigma_{\rm H}$, and write 
\begin{equation}
    \zeta_{\rm A, sec} = \frac{\sigma_{\rm A}}{\sigma_{\rm H}} \zeta_{\rm H, sec}.
\end{equation}
As an example, for C, $\sigma_{\rm C}/\sigma_{\rm H} = 3.92$, hence the volumetric production rate can be written as
\begin{equation}
\begin{split}
    \left(\frac{{\rm d}n_{\rm C^+}}{{\rm d}t}\right)_{\rm sec} &= 3.92n_{\rm C} \frac{\langle\phi_{\rm H,0}\rangle}{1+1.82 (n_{\rm H_2}/n_{\rm H})} (\zeta_{\rm H} + \frac{n_{\rm He}}{n_{\rm H}}\zeta_{\rm He})\\
    &= n_{\rm C} \langle\phi_{\rm H,0}\rangle \frac{n_{\rm H}\zeta_{\rm H} + n_{\rm He}\zeta_{\rm He}}{n_{\rm H}+1.82n_{\rm H_2}},
\end{split}
\end{equation}
where $\langle\phi_{\rm H,0}\rangle$ represents an average over the X-ray spectrum.

\subsection{Heating}
\begin{figure}
    \centering
    \includegraphics[width=\columnwidth]{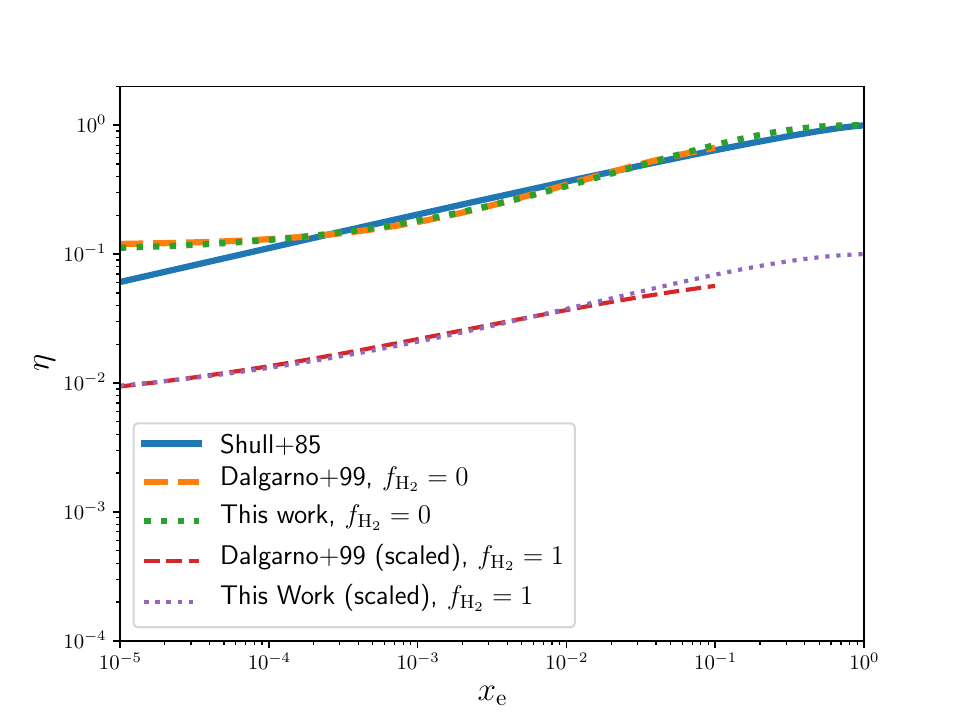}
    \caption{Same as Fig.~\ref{fig:fitsH}, but showing here the heating efficiency from secondary ionisations by X-rays. The $r=1$ cases here have been scaled down by a factor 0.1 only for visualisation purposes.}
    \label{fig:eta}
\end{figure}

While part of the primary electron energy contributes to secondary ionisations and excitations, the rest is deposited as heat. Similarly to what we did for ionisations, we can rewrite the heating rate from electron-atom collisions as \begin{equation}
\begin{split}
    \Gamma_{\rm sec} &= \eta n_{\rm H_{tot}} \int_{\rm E_{min}}^{\rm E_{max}} \bar{\sigma} \frac{J(E)}{E}(E-E_{\rm th}){\rm d}E \\
    &= \eta \sum_i \int_{\rm E_{min}}^{\rm E_{max}} n_i\sigma_i \frac{J(E)}{E}(E-E_{\rm th}){\rm d}E\\
    &=\eta \sum_i \Gamma_{i},
\end{split}
\end{equation} 
where $\eta$ is the heating efficiency for the considered element mixture, $\Gamma_i = n_i H_i$, with $H_i$ the normalised heating rate, which can be approximated as  
\begin{equation}
    H_i \simeq \int_{E_{\rm min}}^{E_{\rm max}} \sigma_i(E) J(E)d E
\end{equation}
since, in the X-ray band, $E\gg E_{\rm th}$.
Also in this case, we follow the formalism by S85. However, instead of using their Eq.~1, we assume that after $N$ collisions the electron will come at rest, hence its energy will be negligible. Since both secondary ionisations and excitations go to zero for $x_{\rm e}\rightarrow 1$, the heating efficiency should also reach unity. Hence, we define $\eta_i \simeq 1 - C*(1-x_{\rm e}^a)^b$, and we fit the results by D99, getting 
\begin{eqnarray}
     \eta_{\rm HeH} &=& 1 - 0.8957(1-x_{\rm e}^{0.5078})^{2.9620}\\
    \eta_{\rm H_2He} &=& 1 - 0.9377(1-x_{\rm e,eff}^{0.3428})^{1.4169}.\\
\end{eqnarray}
The total efficiency can then be written as
\begin{equation}
    \eta = \frac{10r\eta_{\rm H_2He} + \eta_{\rm HeH}}{10r+1},
\end{equation}
where $r=n_{\rm H_2}/n_{\rm H}$.
Our new fit is reported in Fig.~\ref{fig:eta} for $r=0$ and $r=1$, where the $r=1$ curves have been scaled down by a factor 0.1 only to better distinguish the two cases, that would have otherwise overlapped significantly. Compared to the original S85 results, our new fits agree very well with those by D99, and again naturally extend to higher $x_{\rm e}$ in a similar way to S85.  

\section{Feedback events}
\label{app:feedback}
\begin{figure}
\centering
    \includegraphics[width=\columnwidth]{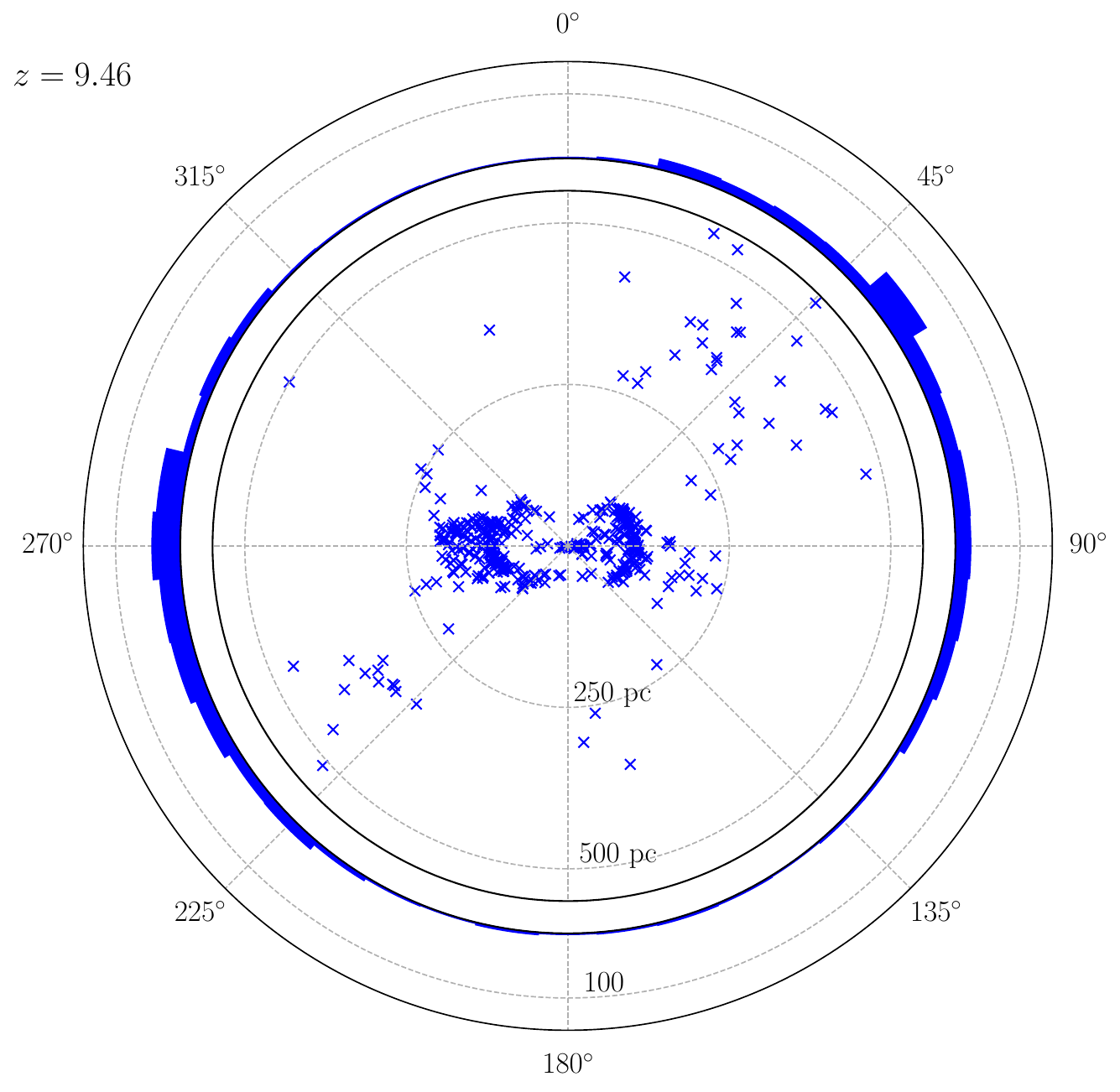} 
    \includegraphics[width=\columnwidth]{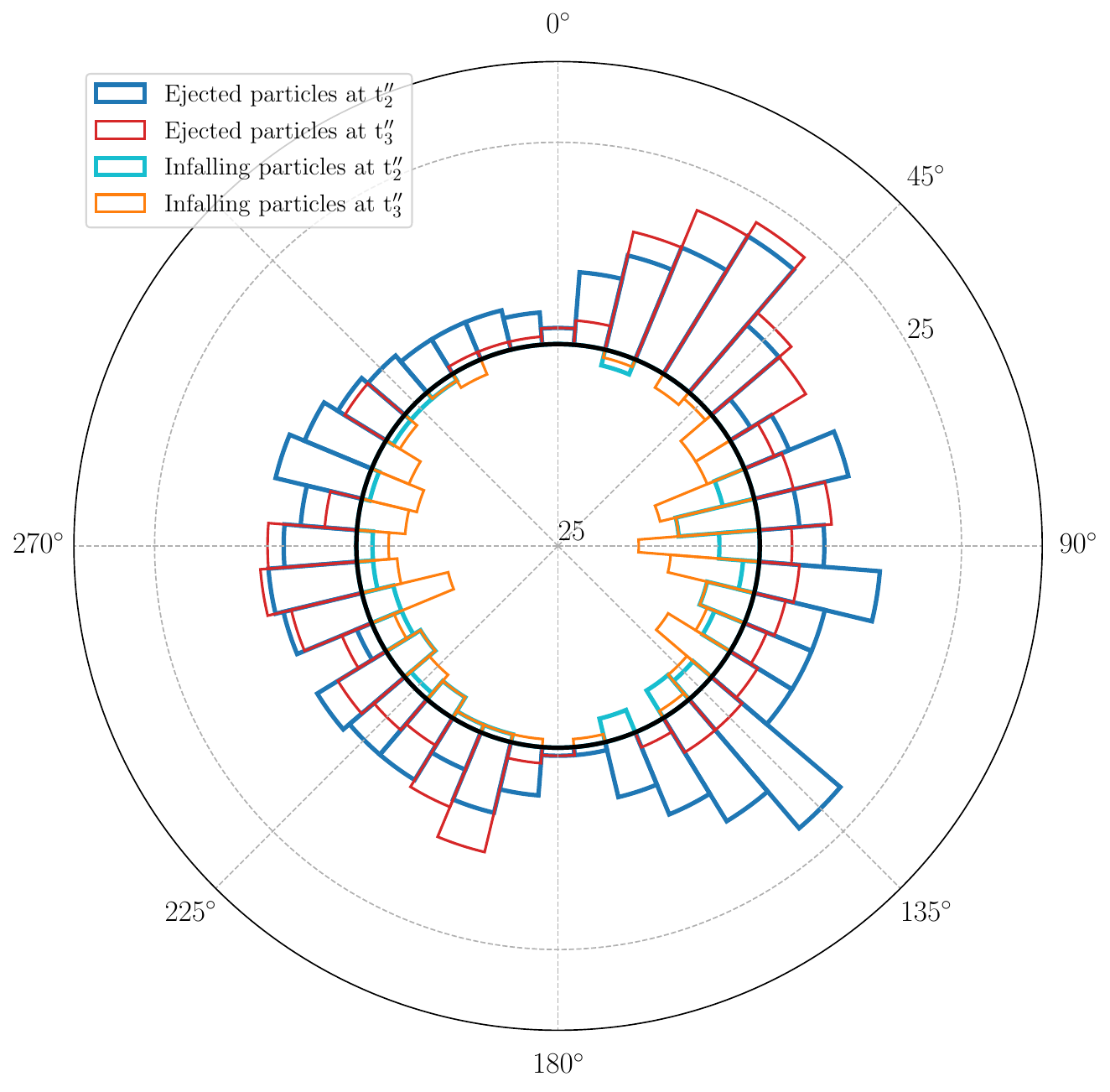} 
\caption{Same analysis as in Fig.~\ref{fig:ang_distr} but for $z \sim 9.46$.}
\label{fig:polar_appendix}
\end{figure}

To support the hypothesis that the cavity observed in our simulation is the result of continuous energy injection in the central region, rather than a single energetic event, we present here the same analysis shown in Fig.~\ref{fig:ang_distr}, but at redshift $z \sim 9.46$. At this stage, the MBH has just entered the super-Eddington accretion regime and begins to grow efficiently, as illustrated in Fig.~\ref{fig:hmad_evol}. The galactic disc is relatively thick, and the accreting material comes from different directions (top panel of Fig.~\ref{fig:polar_appendix}). The ejected material, as shown in the bottom panel, leaves the galactic centre almost isotropically, but escapes the galaxy only perpendicular to the galactic disc, as the outflow shocking with the high-density gas in the disc is immediately slowed down and falls back onto the MBH within 5~Myr.
    
\end{appendix}

\end{document}